\documentclass[11pt,a4paper]{article}

\usepackage{jcappub}

\usepackage{ragged2e}
\usepackage{graphicx}

\usepackage[utf8]{inputenc}
\usepackage{amsmath,amsfonts,amssymb}
\usepackage{color}
\usepackage{soul}

\makeatletter
\gdef\@fpheader{}
\makeatother

\def\fnl{f_{_{\rm NL}}}
\def\d{{\rm d}}
\def\pp{p_{\phi}}
\def\dpp{\delta p_{\phi}}
\def\dph{\delta\phi}

\def\Mpc{{\rm Mpc}}

\begin{document}

\title{Estimation of imprints of the bounce in loop quantum cosmology on the bispectra of cosmic microwave background}
\author{Roshna K}
\emailAdd{roshnak.217ph005@nitk.edu.in }
\author{ and V. Sreenath}
\emailAdd{sreenath@nitk.edu.in}
\affiliation{Department of Physics, National Institute of Technology Karnataka, Surathkal, Mangaluru 575025, India.}


\abstract{
Primordial non-Gaussianity has set strong constraints on models of the early universe. 
Studies have shown that Loop Quantum Cosmology (LQC), which is an attempt to extend inflationary scenario to planck scales, leads to a strongly scale dependent and oscillatory non-Gaussianity. In particular, the non-Gaussianity function $\fnl (k_1,\, k_2,\, k_3)$ generated in LQC, though similar to that generated during slow roll inflation at small scales, is highly scale dependent and oscillatory at long wavelengths. In this work, we investigate the imprints of such a primordial bispectrum in the bispectrum of Cosmic Microwave Background (CMB). 
Inspired by earlier works, we propose an analytical template for the primordial bispectrum in LQC.
We write the template as a sum of strongly scale dependent and oscillatory part, which captures the contribution due to the bounce, and a part which captures the scale invariant behaviour similar to that of slow roll. We then compute the reduced bispectra  of temperature and electric polarisation and their three-point cross-correlations corresponding to these two parts. 
We show that the contribution from the bounce to the reduced bispectrum is negligible compared to that from the scale-independent part. Thus, we conclude that the CMB bispectra generated in LQC will be similar to that generated in slow roll inflation. We conclude with a discussion of our results and its implications to LQC.
}

\maketitle
\section{Introduction}
Numerous theoretical insights together with several observational efforts, spanned over a century, have enabled us to arrive at a compelling model of our Universe referred to as the standard model or the Lambda Cold Dark Matter ($\Lambda$CDM) model \cite{Aghanim:2018eyx}. According to this model, the seeds of the current distribution of galaxies spread over the fabric of spacetime known as the large scale structure were sown during the earliest phase of the universe. Tiny perturbations generated in the early universe lead to tiny anisotropies in the Cosmic Microwave Background (CMB) which in turn lead to the inhomegeneous large scale distribution of galaxies that we see today. Though we have a good level of understanding of this evolution, several details are yet to be worked out. One such detail concerns the origin of these perturbations in our Universe. 
\par
Inflation, see, for instance, \cite{Riotto:2002yw, Bassett:2005xm, Sriramkumar:2009kg, Baumann:2009ds}, due to its simplicity, provides the most popular explanation for the origin of these perturbations \cite{Akrami:2018odb, Martin:2013nzq} (For a discussion on alternate views, see \cite{Steinhardetal, ResponsetoSteinhardetal}.). 
In inflationary scenario, quantum fluctuations in the inflaton  leads to the primordial perturbations. Appealing to the nearly de Sitter symmetry of the spacetime during inflation, we assume that at a time when the perturbations are sufficiently sub-horizon, quantum perturbations are generated in the Bunch-Davies vacuum. Such a prescription has been highly successful, in that, it leads to primordial perturbations that are nearly Gaussian and scale invariant as demanded by observations \cite{Aghanim:2018eyx, Akrami:2018odb, Akrami:2019izv}. 
Moreover, one of the earliest and well motivated model, namely, the Starobinsky model \cite{Starobinsky:1980te} is well favoured by data from Planck.
Even though inflation is successful, it is still an incomplete theory. We do not take in to account the evolution of perturbations before the time at which the initial conditions are imposed. In fact, inflation does not account for the physics in the planck regime close to the big bang singularity. There have been several attempts to address these issues. In this work, we will concern ourselves with loop quantum cosmology (LQC) \cite{Ashtekar:2006rx,Ashtekar:2006wn,Ashtekar:2011ni,Agullo:2016tjh,Agullo:2013dla}. 
\par
Loop quantum cosmology is an attempt to extend inflationary scenario to the planck regime using principles of loop quantum gravity \cite{Bojowald:2001xe,Ashtekar:2003hd,MenaMarugan:2011va,Banerjee:2011qu,Ashtekar:2011ni,Agullo:2016tjh,Agullo:2013dla}. In LQC, quantum gravitational effects in the planck regime leads to a quantum bounce \cite{Ashtekar:2006rx,Ashtekar:2006wn}. Thus in LQC, a quantum bounce precedes the inflationary phase. Generation and evolution of perturbations in LQC have been extensively studied at the level of primordial power spectra \cite{Bojowald:2008jv,Bojowald:2010me,Agullo:2012sh,Agullo:2012fc,Agullo:2013ai, Fernandez-Mendez:2013jqa,Fernandez-Mendez:2014raa,Barrau:2014maa,deBlas:2016puz,Agullo:2015tca,Agullo:2016hap,Ashtekar:2016wpi,Martinez:2016hmn, Gomar:2017yww,Zhu:2017jew,Agullo:2018wbf,Li:2019qzr,ElizagaNavascues:2020fai,Li:2020mfi,Agullo:2020wur,Agullo:2020iqv,Ashtekar:2020gec,ElizagaNavascues:2020uyf,Martin-Benito:2021szh,Li:2021mop, Ashtekar:2021izi, Agullo:2021oqk} and primordial non-Gaussianity \cite{Agullo:2017eyh,Agullo:2015aba,Sreenath:2019uuo,Zhu:2017onp}. In general, studies show that the effect of the bounce is to introduce an additional scale corresponding to the curvature at the bounce. Modes of perturbations which have comparable length to this new scale gets modified leading to a highly scale dependent power spectrum. At smaller wavelengths, the perturbations are not affected by the bounce and the power spectrum is nearly scale invariant as in slow roll inflation \cite{Agullo:2015tca}. Perturbations show a similar behaviour at second order in perturbations \cite{Agullo:2017eyh,Sreenath:2019uuo}. Studies show that primordial non-Gaussianity quantified using the function $\fnl (k_1,\, k_2,\, k_3)$, at scales comparable to the curvature at the bounce, is strongly scale dependent and oscillatory with a very large amplitude. At smaller scales, the $\fnl (k_1,\, k_2,\, k_3)$ is similar to that in slow roll. Studies also show that the effects of the bounce are visible in the bispectrum at smaller wavelengths than in the case of power spectrum. In other words, bispectrum is scale dependent and oscillatory even at wavelengths at which power spectrum is nearly scale invariant. In this sense, bispectrum is more sensitive to the bounce than the power spectrum. 
\par
Assuming sixty or so e-folds of inflation, the scale at which the imprints of the bounce, on primordial perturbations, occur depends on the amount of expansion between the bounce and the onset of inflation. Observational constraints from the CMB temperature power spectrum demand that any departure from scale invariance should happen only at multipoles of $\ell \lesssim 30$ \cite{Aghanim:2018eyx, Aghanim:2019ame}. 
If we assume that, the effects of primordial power spectrum on the CMB is observable at $\ell \lesssim 30$, then, since the bispectrum is more sensitive to the effects of the bounce than the power spectrum \cite{Agullo:2017eyh,Sreenath:2019uuo}, there is a possibility that the imprints of large, scale dependent and oscillatory primordial non-Gaussianity is observable at larger multipoles. Hence it is important to investigate the consistency of LQC with observations by Planck. With this motivation, in this work, we compute the imprints of such a non-Gaussianity in the temperature (T) and electric polarisation (E) of the CMB. We assume an analytical template for primordial non-Gaussianity generated in LQC, compute the $\langle TTT \rangle$, $\langle TTE\rangle$, $\langle TEE \rangle$ and $\langle EEE \rangle$ correlations and show that they are similar to those generated in slow roll inflation and hence is consistent with observations by Planck. 
\par
The rest of the paper is organised as follows. In the next section, we briefly introduce the essentials of LQC and present analytical templates for the primordial power spectrum and bispectrum. 
In section \ref{sec:cmb}, we discuss the essential formulae to compute the three-point correlation functions of anisotropies in temperature and electric polarisation. In section \ref{sec:lqcbispectra}, we apply these formulae to LQC. We present the numerical techniques and our calculation of reduced bispectra of temperature fluctuations  and electric polarisation and their three-point cross-correlations in section \ref{sec:numerics}. We conclude the paper with a summary and discussion of our results and their consequences to LQC in section \ref{sec:discussion}.
\section{Loop quantum cosmology\label{sec:lqc}}
In this section, we will discuss the essentials of LQC that is relevant to this paper (for reviews, see, for instance, \cite{Ashtekar:2011ni,Agullo:2016tjh,Agullo:2013dla}). In particular, we will discuss LQC as applied to FLRW geometries sourced by a scalar field $\phi$ and scalar perturbations $\delta\phi(\vec x)$ living on this background. 
\subsection{Background}\label{sec:background}
\begin{figure}
\centering
\begin{tabular}{cc}
\includegraphics[width=0.5\linewidth]{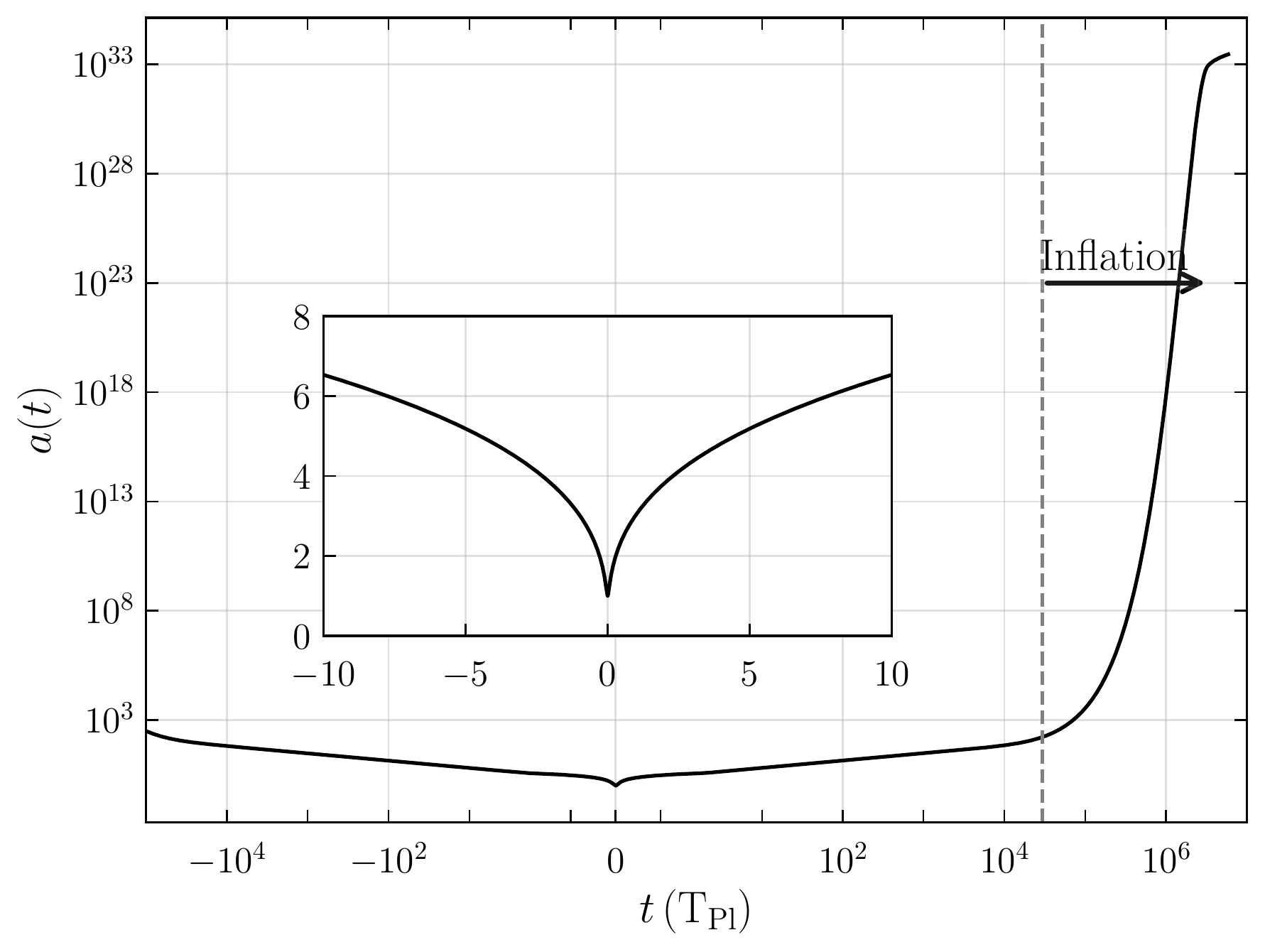}&
\includegraphics[width=0.5\linewidth]{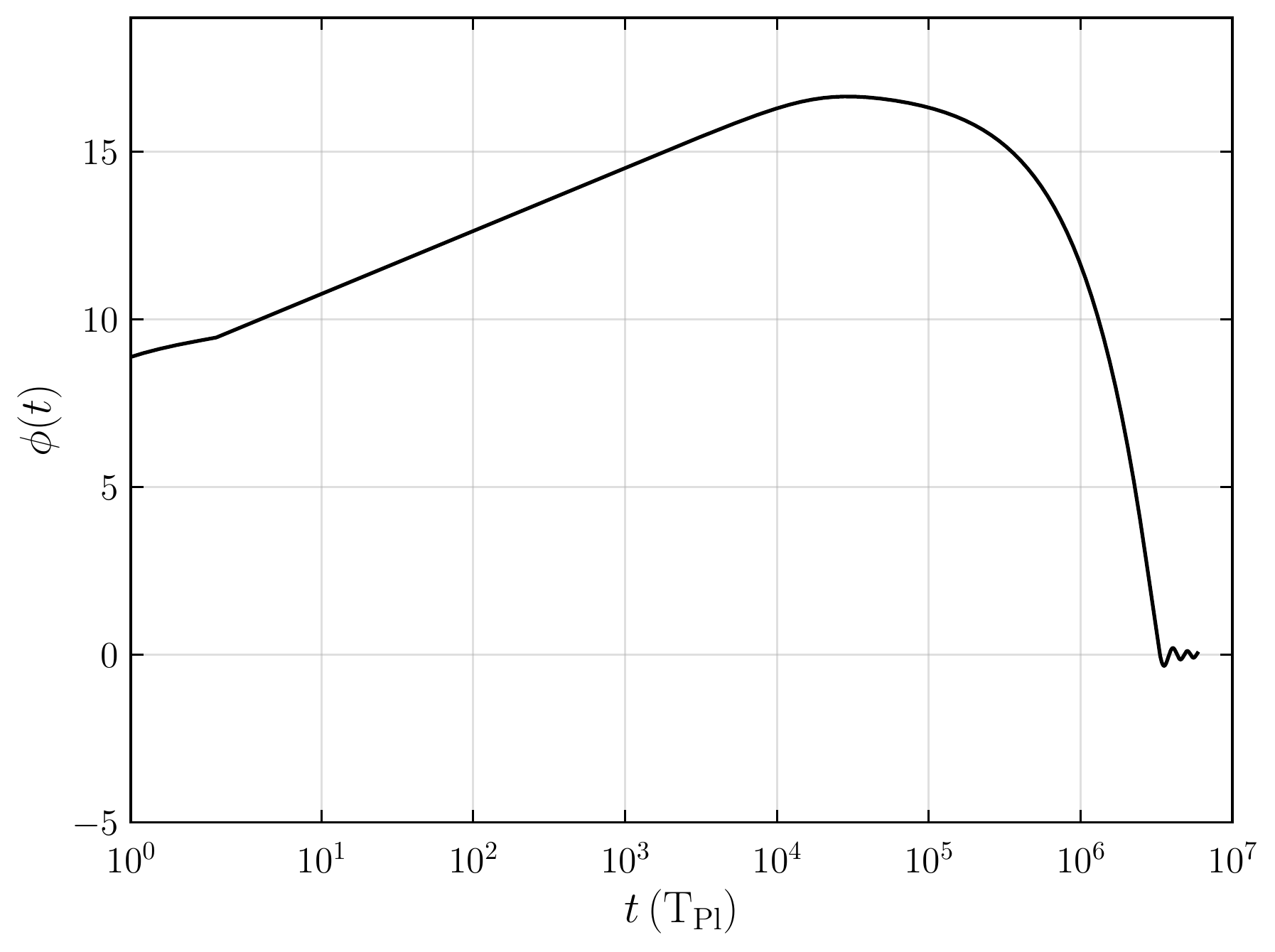}
\end{tabular}
\caption{\label{fig:background} Figure illustrates the behaviour of scale factor (left) and scalar field governed by a quadratic potential (right) in LQC. As mentioned in the text, scale factor undergoes a bounce preceding inflation. Scalar field starts rolling up the potential from its value at the bounce until its kinetic energy becomes zero and then starts slowly rolling down the potential leading to inflation. In obtaining this plot, we have set mass of the scalar field, $m = 6.39\times10^{-6} \rm M_{Pl}$.}
\end{figure}

In LQC, FLRW background geometry is described by a wavefunction $\Psi_{_{\rm FLRW}}(v,\,\phi)$, which satisfies the equation $\hat{\cal H}_{_{\rm FLRW}} \Psi_{_{\rm FLRW}}(v,\,\phi)\, =\, 0$, where $\hat{\cal H}_{_{\rm FLRW}}$ is the Hamiltonian operator corresponding to the classical background Hamiltonian and $v$ is the volume factor which is proportional to the cube of scale factor $a$. Numerical investigations of such a system has shown that the scale factor undergoes a bounce \cite{Ashtekar:2006rx,Ashtekar:2006wn, Diener:2013uka, Diener:2014mia}. 
It turns out, if the wave function is sharply peaked over the values of scale factor, the behaviour of scale factor can be described by certain effective equations \cite{Ashtekar:2006rx,Ashtekar:2006wn, Diener:2013uka, Diener:2014mia, Agullo:2016hap}, namely
\begin{subequations}\label{eqn:modfriedmann}
\begin{eqnarray}
\left(\frac{\dot a}{a}\right)^2\, &=&\, \frac{\kappa}{3}\rho\, \left( 1\, -\, \frac{\rho}{\rho_{sup}} \right),\\
\frac{\ddot a}{a}\, &=&\, -\frac{\kappa}{6}\,\rho\,\left( 1\, -\, 4\,\frac{\rho}{\rho_{sup}} \right)\, -\, \frac{\kappa}{2}\, P\,\left( 1\,-\,2\,\frac{\rho}{\rho_{sup}} \right),
\end{eqnarray}
\end{subequations}
where $\rho$, $P$ are the energy density and pressure of the scalar field and $\kappa\, =\, 8\,\pi\,G$.
From the above expression, it is clear that at $\rho\,=\,\rho_{sup}$, Hubble parameter $H\,=\dot a/a\,=\,0$ and  $\ddot a/a > 0$ {\it i.e.~} scale factor is at minimum. In other words, the universe undergoes a bounce at $\rho\,=\,\rho_{sup}$. Computation of black hole entropy pegs the value of $\rho_{sup}$ to be equal to $0.41 m_{_{\rm Pl}}^4$ \cite{Meissner:2004ju}. Further, if we assume that the scalar field is governed by a potential $V(\phi)$, then the evolution of scalar field is given by 
\begin{equation}\label{eqn:eomphi}
\ddot\phi\, +\, 3\,H\,\dot\phi\, +\,V_\phi\,=\,0,
\end{equation}
where $V_\phi\,=\,{\rm d}V/{\rm d}\phi$. For a suitable potential, inflationary phase will set in after the bounce \cite{Ashtekar:2009mm,Ashtekar:2011rm,Bolliet:2017czc,Bonga:2015kaa,Bonga:2015xna,Zhu:2017jew}. 
The energy density of LQC near the bounce is dominated by the kinetic energy of the scalar field, or in other words, the potential energy is negligible compared to the kinetic energy at the bounce. Hence, to understand the imprints of the bounce, it is sufficient to work with a simple model such as a quadratic potential, see, for instance, \cite{Zhu:2016dkn}. 
The background dynamics in LQC with a scalar field governed by a quadratic potential is illustrated in Figure \ref{fig:background}.
We set the mass of the scalar field in such a way that the amplitude of the power spectrum at larger wavenumbers is consistent with the observation by Planck. 
Moreover, we have set the value of scalar field at the bounce to be $5.6\,{\rm M_{Pl}}$. 

\subsection{Perturbations}\label{sec:Pert}
We will follow dressed metric approach to describe primordial perturbations in LQC \cite{Agullo:2012sh,Agullo:2012fc,Agullo:2013ai,Agullo:2015tca, Agullo:2017eyh, Sreenath:2019uuo}. In this approach, we assume that the wavefunction takes the form $\Psi\,=\,\Psi_{_{\rm FLRW}}(v,\,\phi)\,\otimes\, \delta\Psi(v,\,\phi,\,\delta\phi)$, which satisfies the equation $\hat {\cal H} \Psi\, =\, 0$, where $\hat {\cal H}\, = \hat{\cal H}_{_{\rm FLRW}}\, + \hat{\cal H}_{{\rm pert}}$. As mentioned earlier, $\Psi_{_{\rm FLRW}}(v,\,\phi)$ satisfies the equation $\hat{\cal H}_{_{\rm FLRW}} \Psi_{_{\rm FLRW}}(v,\,\phi)\, =\, 0$. Perturbations are treated as test fields living on the background FLRW geometries described by $\Psi_{_{\rm FLRW}}(v,\,\phi)$. In practice, this implies that perturbations can be evolved using the classical Hamiltonian but with the background functions in them described by the effective equations. This is similar to perturbations living as test fields on a curved space time described by a `dressed' metric which satisfies the effective equations. In order to compute primordial bispectrum, we need to consider Hamiltonian up to third order in perturbations, {\it i.e.~} we need ${\cal H}_{{\rm pert}}\, =\, {\cal H}^{(2)}\,+\, {\cal H}^{(3)}$. There are two approaches to arrive at the Hamiltonian describing perturbations, one can either use gauge invariant variables or rather work with a fixed  gauge. We follow the latter approach. In particular, we will work with spatially flat gauge \cite{Maldacena:2002vr,Agullo:2017eyh}. 
\par
The second order Hamiltonian describing perturbations $\delta\phi$ in the spatially flat gauge is 
\begin{eqnarray} \label{eqn:H2}
 \mathcal{H}^{(2)}\, =\,\int \d^3 x \, N \, \, \mathbb{S}^{(2)}(\vec x)=\, N\frac{1}{2}\,\int \d^3 x \,  \biggl[\, \frac{1}{\,a^3}\, \delta p\phi^2\, +\, a^3\, (\partial \delta\phi)^2\, 
 +\,a^{3}\,  {\cal U}\, \dph^2\biggr]\, ,
\end{eqnarray}
with  the potential  ${\cal U}$ given by
\begin{equation} \label{eqn:U} 
{\cal U}=-9 \frac{\pp^4}{a^8\pi_a^2}+\frac{3}{2}\kappa\frac{\pp^2}{a^6}-\frac{6 \, \pp }{a\, \pi_a} V_{\phi}+ V_{\phi \phi} +6\frac{\pp \dot{p}_{\phi}}{a^4\, \pi_a}-3\frac{\pp^2 \, \dot{\pi}_a}{a^4\, \pi_a^2}-3\, \frac{\dot a\, \pp^2}{a^5\, \pi_a} \, .
\end{equation}
In the above expressions, $\pi_a$, $p_\phi$ and $\dpp$ are momenta conjugate to $a$, $\phi$ and $\delta\phi$ respectively. Setting lapse $N\, =\,1$ will imply cosmic time and $N\,=\,a$ corresponds to conformal time. 
Hamiltonian at third order in perturbations is
\begin{eqnarray}\label{eqn:H3}
 \mathcal{H}^{(3)}\, 
 &=&
 \, N\,\int\, \d^3  x\, \biggl[ 
\left(  \frac{9\,\kappa\,p_{\phi}^3}{4\,a^4\,\pi_a}-\frac{27 \, \pp^5}{2\, a^6\pi_a^3} -\, \frac{3\,a^2\,p_{\phi}\,V_{\phi\phi}}{2\,\pi_a}\, 
 +\frac{a^3\,V_{\phi\phi\phi}}{6} \right) \,\delta\phi^3\, \nonumber \\
 &&-\, \frac{3\,p_{\phi}}{2\,a^4\,\pi_a}\,\delta p_{\phi}^2\, \delta\phi\,-\frac{9 \, \pp^3}{ a^5 \pi_a^2} \, \dpp\dph^2
-\, \frac{3\,a^2\, p_{\phi}}{2\, \pi_a}\delta\phi\, (\vec{\partial}\delta\phi)^2 +\, \frac{3\,p_{\phi}^2}{N\, a\,\pi_a}\,\delta\phi^2 \partial^2\chi\, +\, \frac{3}{2}\frac{a^2\,p_{\phi}}{N^2\,\kappa\,\pi_a}\,\delta\phi\,\partial^2\chi\,\partial^2\chi \nonumber \\
 &&+\,3\, \frac{\pp^2}{N\, a\,\pi_a}\, \delta\phi\, \partial^i\chi\partial_i\delta \phi +\frac{1}{N}\, \delta p_{\phi}\,\partial_i \delta\phi\, \partial^i\chi\,
 -\, \frac{3}{2}\frac{a^2\,p_{\phi}}{N^2\,\kappa\,\pi_a}\, \delta\phi\, \partial_i\partial_j\chi\, \partial^i\partial^j\chi\,
 \biggr],
\end{eqnarray}
where $\partial^2\chi\, =\,\left(-3\, N\, \kappa/a\right) \,\biggl[ \biggl(\,\frac{\pp}{2} -\, \frac{a^5\, V_{\phi}}{\kappa\,\pi_a} \biggr)\delta \phi\, -\, \frac{\pp}{\kappa\, a\, \pi_a}\delta p_{\phi}\, \biggr]$.
\par
From the second order Hamiltonian ${\cal H}^{(2)}$, one can derive the free evolution of the scalar perturbation, given by,
\begin{equation}\label{eqn:eom}
\left(\Box\, -\, {\cal U}(t)\right)\delta\phi(\vec x,\,t)\, =\, 0,
\end{equation}
where $\Box$ is the d'Alembertian of the FLRW background metric. The third order Hamiltonian ${\cal H}^{(3)}$ provides the self-interaction of the scalar perturbations. 
\par
The perturbations, since they evolve through the bounce and then through the inflationary phase, carry signatures of the early universe which they imprint on the CMB. 
Perturbations are quantified using correlation functions. In order to compute  correlation functions, one need to promote $\delta \phi$ to an operator $\hat{\delta\phi}$. The field operator $\hat{\delta\phi}$ is then expanded in terms of annihilation and creation operators as 
\begin{eqnarray}
\hat{\delta\phi}({\vec x},\eta) = \int\frac{{\rm d}^3 k}{(2\pi)^3} \,  \hat{\delta\phi}_{\vec{k}}( \eta)\, e^{i{\vec k}\cdot{\vec x}}= \int\frac{{\rm d}^3 k}{(2\pi)^3} \left(\hat A_{\vec k}~\varphi_k(\eta) + \hat A^\dagger_{-\vec k}
~\varphi_{k}^*(\eta)\right) e^{i{\vec k}\cdot{\vec x}}
\end{eqnarray}
where $[\hat A_{\vec k},\hat A^{\dagger}_{\vec k'}]=\hbar \, (2\pi)^3\, \delta^{{(3)}}(\vec k+\vec k')$, $[\hat A_{\vec k},\hat A_{\vec k'}]= 0 = [\hat A^{\dagger}_{\vec k},\hat A^{\dagger}_{\vec k'}]$. The mode functions $\varphi_k (\eta)$ satisfy the equation 
\begin{equation}\label{eqn:eomvarphi}
\varphi_k^{\prime\prime} + 2\frac{a '}{a} \varphi_k^\prime + (k^2 +
a^2 \, {\cal U})\,  \varphi_k =0\, ,
\end{equation}
where  $k^2\equiv  k_i k_j\, \delta^{ij}$ is the comoving wavenumber, and prime indicates derivative with respect to conformal time. 
The scalar power spectrum of $\hat{\dph}$ is a dimensionless function that quantifies the two-point correlation in momentum space via
\begin{equation} \label{2pc}
  \langle 0|\hat{\dph}_{\vec k}( \eta) \hat{\dph}_{\vec k^\prime}(\eta)|0\rangle \equiv 
(2\pi)^3\delta^{{(3)}}({\vec k}+{\vec k^\prime}) \frac{2\pi^2}{k^3} \mathcal P_{\dph}(k, \eta)\, ,
\end{equation}
where $|0\rangle$ is the vacuum annihilated by the operators $\hat A_{\vec{k}}$ for all $\vec{k}$. 
Power spectrum, in terms of mode functions, is  $\mathcal P_{{\dph}}(k,\eta) = (\hbar\,{k^3}/{2\pi^2})\,|\varphi_k(\eta)|^2$.
\par
The three-point function of $\hat{\delta\phi}$ at tree level is given by \cite {Maldacena:2002vr,Agullo:2017eyh}
\begin{eqnarray}\label{eqn:tree}
	\langle0|\, \hat{\delta\phi}_{\vec k_1}( \eta)\,\hat{\delta\phi}_{\vec k_2}( \eta)\hat{\delta\phi}_{\vec k_3}( \eta)\,|0\rangle &=&\, 
	-\,i/\hbar  \int d \eta' \langle 0|\left[ \hat{\dph}^{\rm I}_{{\vec k}_1}( \eta) \hat{\dph}^{\rm I}_{{\vec k}_2}(\eta) \hat{\dph}^{\rm I}_{{\vec k}_3}( \eta), \hat{\mathcal H}^{\rm I}_{\rm int}( \eta')\right]|0\rangle \nonumber\\
	&&\,+\,\mathcal{O}(\mathcal H^2_{\rm int}),
	\end{eqnarray}
where $\hat{\mathcal H}^{\rm I}_{\rm int}( \eta)$ is the operator corresponding to $\mathcal{H}^{(3)}$ in the interaction picture.

\par
Even though we worked in spatially flat gauge, it is convenient to compute correlation functions in terms of curvature perturbations ${\mathcal R}$. 
This is because, curvature perturbations have a unique property that they stop evolving after they cross the horizon and remain constant till 
they re-enter horizon towards late radiation domination or during early matter domination epoch, saving us a lot of computational time. 
Curvature perturbations are related to perturbations in scalar field through the relation \cite{Maldacena:2002vr,Agullo:2017eyh}
\begin{eqnarray}
 \label{Rtodph} \mathcal{R}(\vec x,\eta )=- \frac{a}{z} \, \dph(\vec x,\eta)+\left[-\frac{3}{2}+3\frac{V_{\phi}\, a^5}{\kappa\, P_\phi\, \pi_a}+\frac{\kappa}{4}\frac{z^2}{a^2}\right]  \left(\frac{a}{z} \, \dph(\vec x, \eta)\right)^2+\cdots \, ,
\end{eqnarray}
where trailing dots indicate terms that lead to subdominant terms in the three-point functions when evaluated towards the end of inflation.
\par
The power spectrum of curvature perturbation is related to that of scalar modes $\hat{\delta\phi}_{\vec k}(\eta)$ through the relation
\begin{equation}\label{eqn:PR}
 {\mathcal P}_{\cal R}(k)\, =\, \biggl( \frac{a(\eta_{f})}{z(\eta_{f})}\biggr)^2\,{\mathcal P}_{\delta\phi}(k,\,\eta),
\end{equation}
where $z\,=\,-6\,p_{\phi}/(\kappa\,\pi_a)$.
\par
The three-point function of curvature perturbation can be obtained in terms of $\hat{\delta\phi}_{\vec k}(\eta)$ by using Eq. (\ref{Rtodph}) as
\begin{eqnarray}\label{eqn:Rtpf}
 & & \langle 0|\hat{\mathcal{R}}_{{\vec k}_1} \hat{\mathcal{R}}_{{\vec k}_2} \hat{\mathcal{R}}_{{\vec k}_3}|0\rangle=\left(-\frac{a}{z}\right)^3  \langle 0|\hat{\dph}_{{\vec k}_1} \hat{\dph}_{{\vec k}_2} \hat{\dph}_{{\vec k}_3}|0\rangle\nonumber \\ &+& \left(-\frac{3}{2}+3\frac{V_{\phi}\, a^5}{\kappa\, \pp\, \pi_a}+\frac{{\kappa}}{4}\frac{z^2}{a^2}\right)\, \left(-\frac{a}{z}\right)^4\, \Big[\int \frac{d^3p}{(2\pi)^3} \, \langle 0|\hat{\dph}_{{\vec k}_1} \hat{\dph}_{{\vec k}_2}  \hat{\dph}_{{\vec p}}\,  \hat{\dph}_{{\vec k}_3-\vec p}|0\rangle + (\vec k_1 \leftrightarrow \vec k_3)+ (\vec k_2 \leftrightarrow \vec k_3) \, \nonumber \\
&+&\cdots\Big] \, .  
\end{eqnarray}
The wavenumbers of three modes in the three-point function are constrained by a Dirac delta function. We define the scalar bispectrum as the three-point function sans Dirac delta function as
\begin{eqnarray}\label{eqn:BispectrumR}
 \langle 0|\hat{\mathcal{R}}_{{\vec k}_1} \hat{\mathcal{R}}_{{\vec k}_2} \hat{\mathcal{R}}_{{\vec k}_3}|0\rangle\equiv  (2\pi)^3\delta^{(3)}(\vec{k}_1+\vec{k}_2+\vec{k}_3) \, B_{\mathcal{R}}(k_1,k_2,k_3) \, .
\end{eqnarray}
The amplitude of bispectrum can be quantified using a dimensionless function $\fnl(k_1,\,k_2\,,k_3)$, akin to the dimensionless power spectrum ${\cal P}_{\cal R}(k)$ that quantifies two-point correlations,  as 
\begin{eqnarray}
\label{eqn:fNL}  f_{_{\rm NL}}(k_1,k_2,k_3) &\equiv&  -\frac{5}{6}
(2\,\pi^2)^{-2} \left(\, \frac{{\cal P_R}(k_1)}{k_1^3}\,\frac{{\cal P_R}(k_2)}{k_2^3}\,+\, \frac{{\cal P_R}(k_2)}{k_2^3}\,\frac{{\cal P_R}(k_3)}{k_3^3}\, +\, \frac{{\cal P_R}(k_3)}{k_3^3}\,\frac{{\cal P_R}(k_1)}{k_1^3} \right)^{-1}\nonumber\\
&&\,\times \, B_{\mathcal{R}}(k_1,k_2,k_3).
\end{eqnarray}

%

\begin{figure} 
\centering
	\begin{tabular}{c}
		\includegraphics[width=0.7\textwidth]{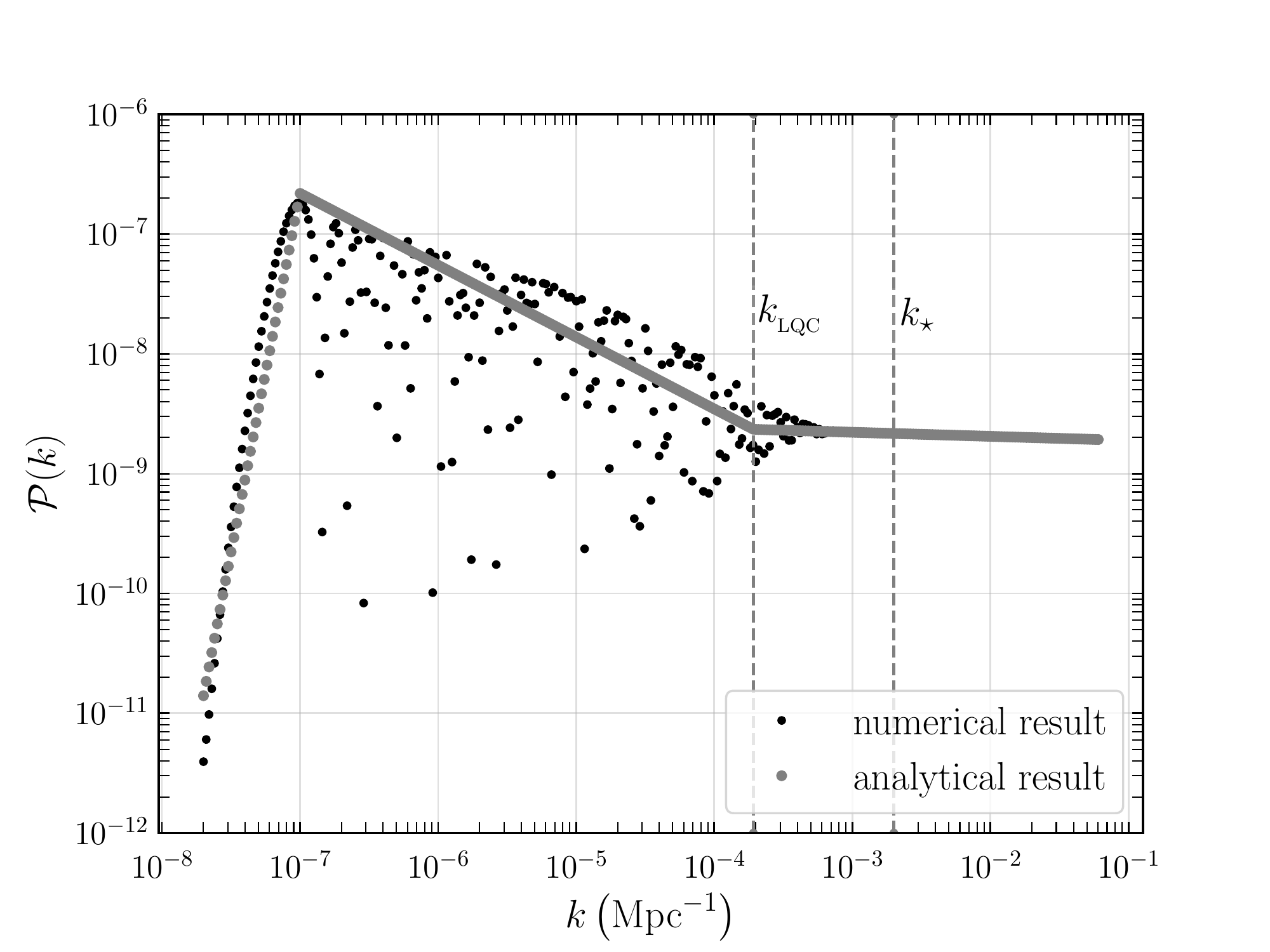}
	\end{tabular}
\caption{\label{fig:Ps} The primordial power spectrum generated in LQC obtained numerically (in black). 
Analytical template for power spectrum given in Eqn. (\ref{eqn:PStemplate}) (in grey). 
}
\end{figure}
\begin{figure} 
\centering
\includegraphics[width=0.7\textwidth]{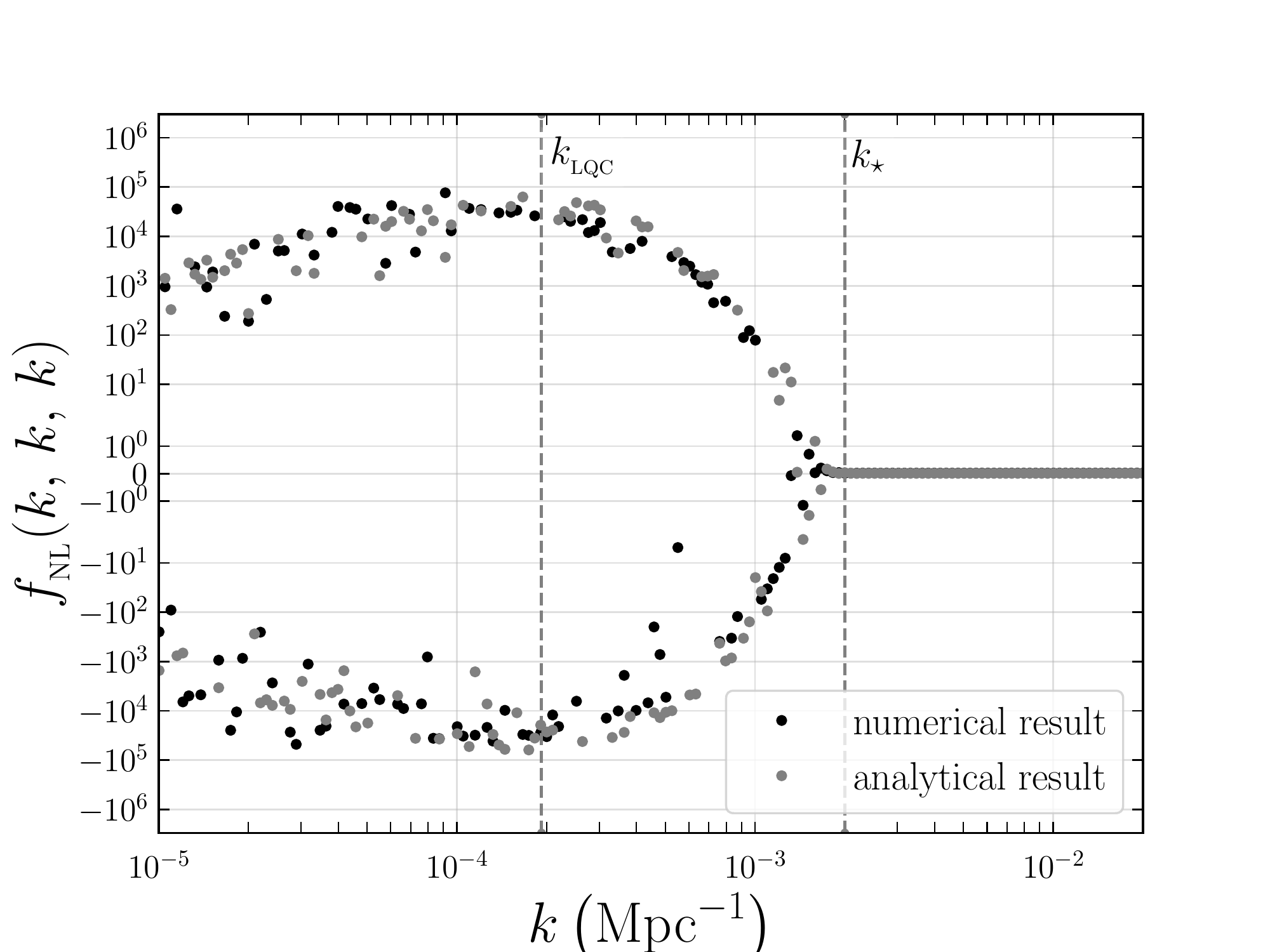}
	\begin{tabular}{cc}
		\includegraphics[width=0.5\textwidth]{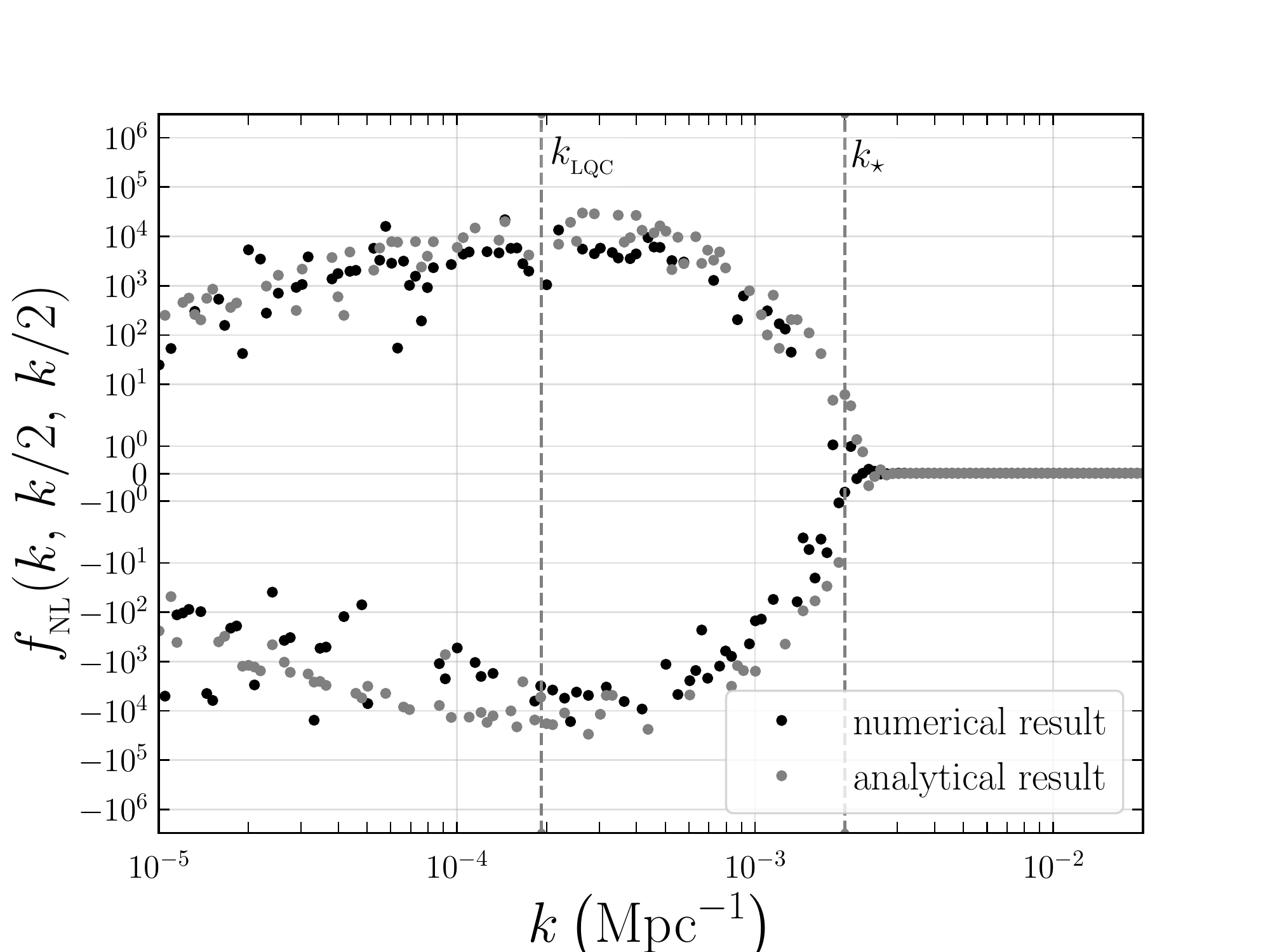}&
		\includegraphics[width=0.5\textwidth]{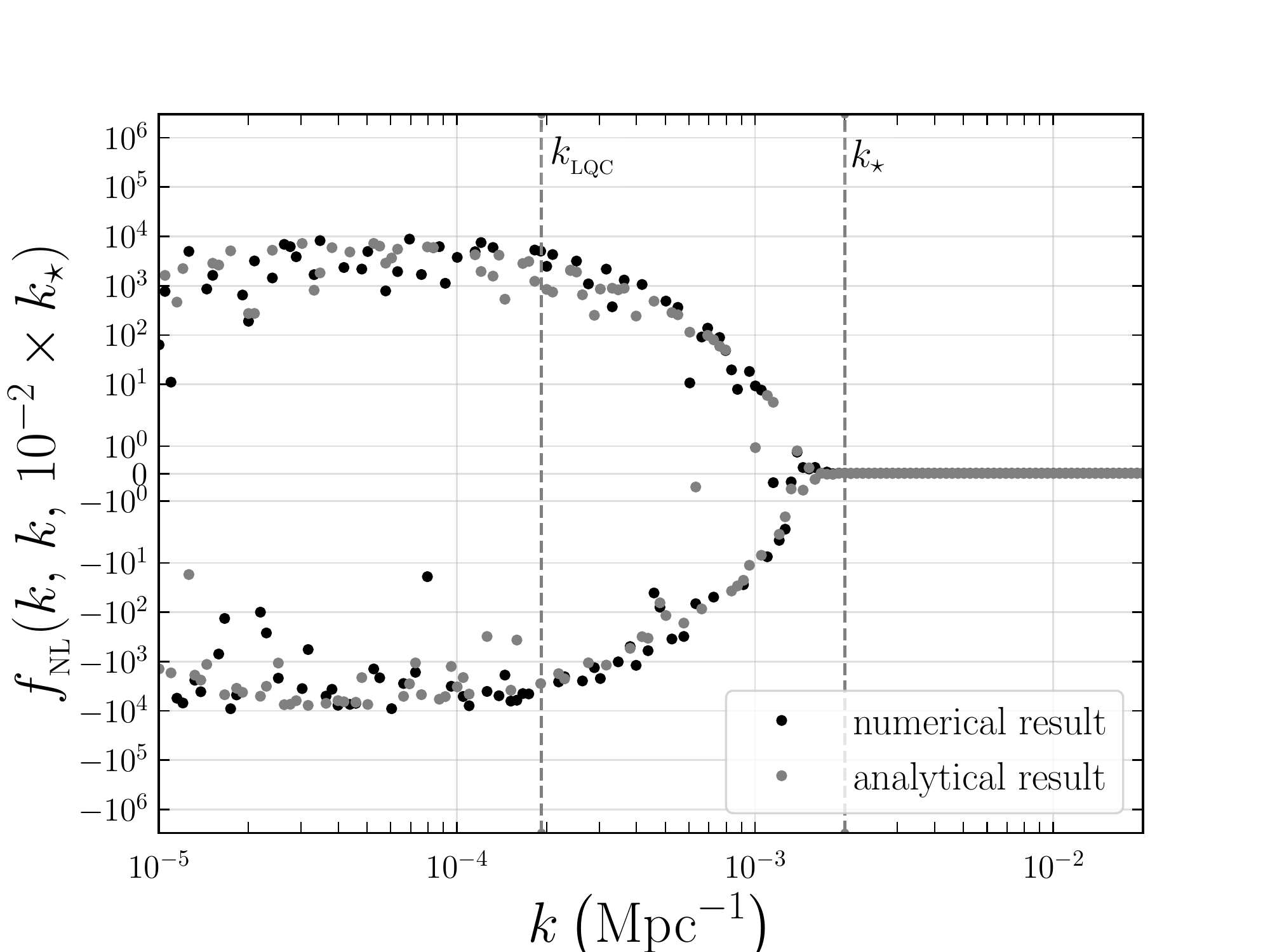}
	\end{tabular}
\caption{\label{fig:fNL} The primordial non-Gaussianity function generated in LQC obtained numerically (in black) for three different configurations {\it viz.} equilateral limit ($k_1 = k_2 = k_3$), folded limit ($k_1 = 2 k_2 = 2 k_3$) and squeezed limit ($k_1 \approx k_2 << k_3$). 
We also plot analytical template for non-Gaussianity corresponding to Eqn.(\ref{eqn:Btemplate}) (in grey).}
\end{figure}
\subsection{Primordial power spectrum and bispectrum\label{sec:template}}
The primordial power spectrum and bispectrum quantifies the primordial perturbations.
They can be calculated numerically. Given the background dynamics described in Eqns.  (\ref{eqn:modfriedmann}, \ref{eqn:eomphi}), the evolution of perturbations are found by solving Eqn. (\ref{eqn:eomvarphi}). The power spectrum of curvature perturbation can then be calculated using Eqn. (\ref{eqn:PR}). 
Calculation of $\langle\,0|\, \hat{\delta\phi}_{\vec k_1}( \eta)\,\hat{\delta\phi}_{\vec k_2}( \eta)\,\hat{\delta\phi}_{\vec k_3}( \eta)\,|0\,\rangle$  requires one to perform integrals in Eqn. (\ref{eqn:tree}).  The $\langle 0|\hat{\mathcal{R}}_{{\vec k}_1} \hat{\mathcal{R}}_{{\vec k}_2} \hat{\mathcal{R}}_{{\vec k}_3}|0\rangle$ three-point function of curvature perturbation can then be calculated using Eqn. (\ref{eqn:Rtpf}). The dimensionless non-Gaussianity function of curvature perturbation is then calculated by using Eqn. (\ref{eqn:fNL}). This numerical calculation of primordial power spectrum and non-Gaussianity has been implemented in \verb|class_lqc| \cite{Agullo:2017eyh, Sreenath:2019uuo}. We present the results obtained using that code in figures  \ref{fig:Ps} and \ref{fig:fNL}. 
From these figures, we see that both power spectrum and $\fnl(k_1,\, k_2,\, k_3)$ are strongly scale dependent at small wavenumbers. We also  see that while the power spectrum generated in LQC is scale dependent for modes $k \lesssim 0.1\,k_\star$, the $\fnl$ is scale dependent for modes $k\lesssim k_\star$, where $k_\star\,=\, 0.002\, \Mpc^{-1}$ is the pivot scale. This means that the effects of the bounce are visible at higher wavenumbers in the bispectrum than in the power spectrum. In this sense, the bispectrum is more sensitive to the bounce than the power spectrum. 
\par
For calculating the three-point functions involving temperature and electric polarisation, one needs to convolve the primordial bispectrum with the CMB transfer functions. For performing this calculation, it is convenient to have analytical templates of primordial power spectrum and bispectrum. In the following two subsections, we attempt to understand the shape of both the spectra and arrive at templates describing their behaviour.

\subsubsection{Analytical template of primordial power spectra}
\begin{figure}
	\includegraphics[width=0.8\textwidth]{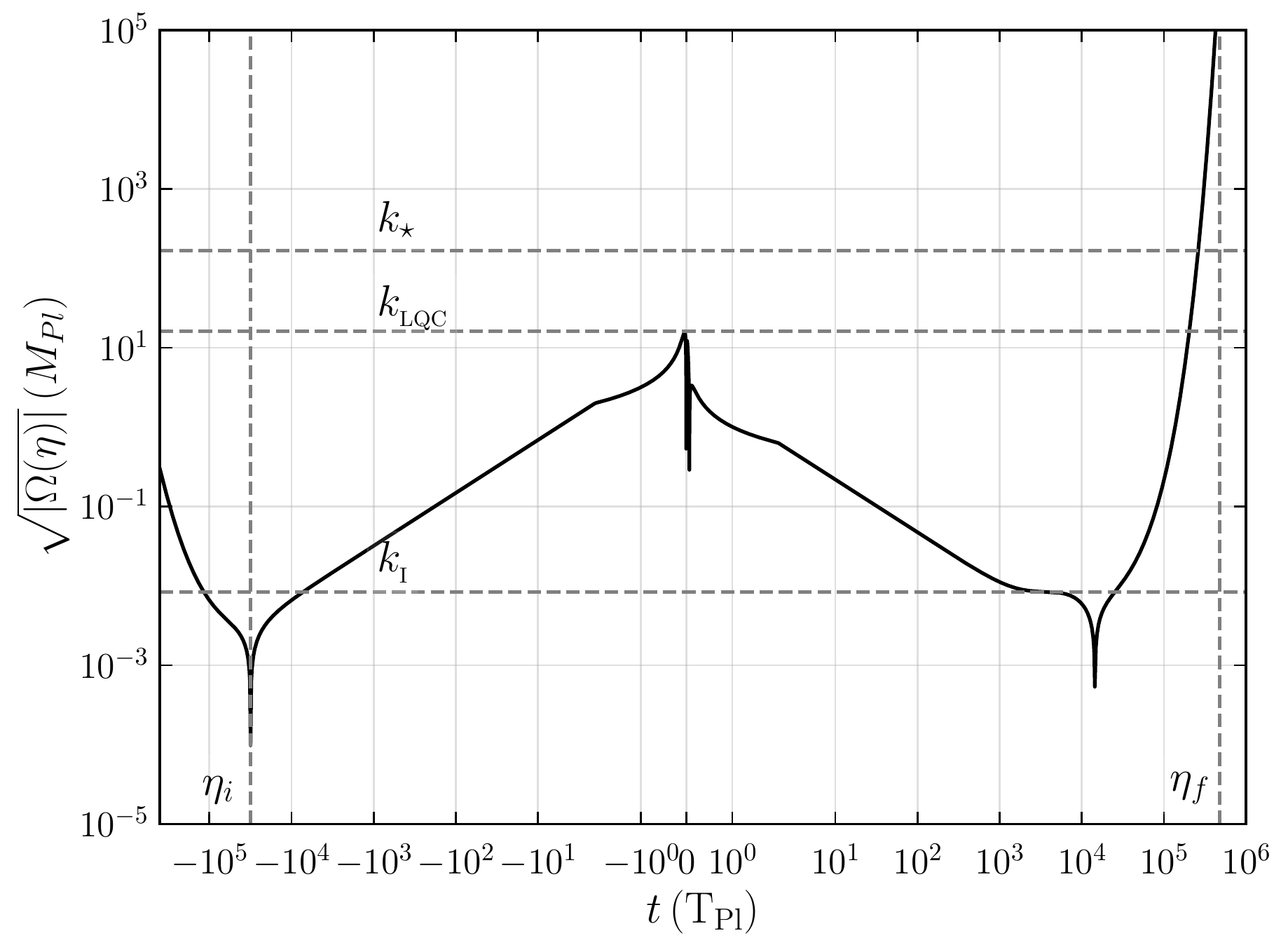}
	\caption{\label{fig:zppbyz} The figure represents the relevant scales in LQC. $k_{_{\rm LQC}}$ is the scale corresponding to the value of curvature at the bounce. $k_{\rm I}$ corresponds to the smallest scale that is sub-horizon during inflation. Adiabatic initial conditions are imposed at $\eta_i$ when all modes of interest are adiabatic and is evolved till a time $\eta_f$ when they are well outside the horizon during inflation. As can be seen, modes $k_{_{\rm LQC}} \gtrsim k > k_{\rm I}$ are excited during the bounce and hence are scale dependent. Modes smaller than $k_{\rm I}$ are mostly superhorizon and hence less excited. Whereas modes with larger wavenumbers, {\it viz.}~$k>>k_{_{\rm LQC}}$, are excited only during horizon crossing towards the end of inflation and hence will be scale invariant.}
\end{figure}
In order to arrive at the template for primordial power spectra, let us rewrite Eqn. (\ref{eqn:eomvarphi}) as
\begin{equation}
 v''_k\, +\, \left(\, k^2\, +\, \Omega(\eta) \right)v_k\,=\,0,
\end{equation}
where $v_k\, =\, a\,\varphi_k$ is the Mukhanov-Sasaki variable and $\Omega(\eta)\,=\,a^2 \, {\cal U}\, -\, \frac{a''}{a}$.
We compare the behaviour of $\sqrt{|\Omega(\eta)|}$ as a function of time with relevant wavenumbers in figure \ref{fig:zppbyz}. As shown in the figure, there exists a time before the bounce where all observationally relevant wavenumbers are adiabatic. Hence we impose adiabatic initial conditions at that time. From the figure, it is also clear that there are two relevant scales in the problem. The value of curvature at the bounce defines a scale $k_{_{\rm LQC}}$ and the value of curvature at the onset of inflation defines a scale $k_{\rm I}$. For the background evolution discussed in \ref{sec:background}, $k_{\rm I} \, =\, 5\,\times\, 10^{-5}\,k_\star$ and $k_{_{\rm LQC}}\,=\, 0.1\, k_\star$.
\par 
The shape of the power spectrum generated in LQC can be understood qualitatively as follows \cite{Agullo:2020fbw, Agullo:2020cvg, Agullo:2021oqk}: (i) The scales which are much larger than $k_{_{\rm LQC}}$, are not effected by the bounce and they will be in Bunch-Davies vacuum at the onset of inflation. This implies that power spectrum of modes $k >> k_{_{\rm LQC}}$ will be nearly scale invariant as in slow roll inflation. 
(ii) Modes which are comparable to $k_{_{\rm LQC}}$ and larger than $k_{\rm I}$ will be excited both during the bounce as well as during the horizon exit during inflation. These modes are in excited non-Gaussian states during the onset of inflation and hence they will be further amplified as they exit the horizon during inflation. Hence, the power spectrum of modes $k_{\rm I}<k<k_{_{\rm LQC}}$ will be strongly scale dependent and amplified. Numerical simulations show that the power has a scale dependence of the form $k^{-0.6}$. (iii) Finally, modes whose wavenumbers are smaller than $k_{\rm I}$ are mostly superhorizon and hence they are less excited. This implies that for very long wavelength, {\it i.e.} infrared, modes the power generated will be smaller. Our simulations show that the power of these  modes scale as $k^6$. 
\par
Putting everything together, we can model the power spectrum of scalar perturbations generated in LQC as  \cite{Agullo:2020fbw, Agullo:2020cvg, Agullo:2021oqk}
\begin{equation}\label{eqn:PStemplate}
\mathcal{P}_{\cal R}(k)=A_s\begin{cases}
		(\frac{k}{k_{_{\rm I}}})^{6}(\frac {k_{_{\rm I}}}{k_{_{\rm LQC}}})^{-0.6}\,\,\,\,\, if \,\,\,\, {k} \leq {k_{_{\rm I}}},\\
		(\frac{k}{k_{_{\rm LQC}}})^{-0.6} \,\,\,\,\,\, if \,\,\,\,{k_{_{\rm I}}}<{k}\leq {k_{_{\rm LQC}}},\\
		(\frac{k}{k_{_{\rm LQC}}})^{(n_s-1)}\,\,\,\, if\,\,\,\, {k}>{k_{_{\rm LQC}}}.
	\end{cases}
\end{equation}
The amplitude of power spectrum $A_s$ and the spectral index $n_s$ have been set to correspond to their values obtained by Planck \cite{Aghanim:2018eyx}. 
The above analytical template for power spectra is drawn along with the exact numerical calculation in figure \ref{fig:Ps}.
\subsubsection{Analytical template of primordial bispectra}
The numerical computation of non-Gaussianity generated in LQC, see figure \ref{fig:fNL}, shows that $f_{_{\rm NL}}(k_1,\, k_2,\, k_3)$ is scale dependent and oscillatory at $k\lesssim\, 10\,k_{_{\rm LQC}}$ and is nearly scale invariant as in the case of slow roll inflation at larger wavenumbers. The exponential decay in the value of $f_{_{\rm NL}}(k_1,\, k_2,\, k_3)$ at $k \gtrsim k_{_{\rm LQC}}$ was explained in \cite{Agullo:2017eyh, Sreenath:2019uuo, Agullo:2021oqk} by analysing the poles of the integrand in Eqn. (\ref{eqn:tree}). We revisit this calculation and provide an improved template for $f_{_{\rm NL}}(k_1,\, k_2,\, k_3)$. 
\par
To compute primordial bispectrum, we need to integrate perturbations from a time $\eta_i$ when adiabatic initial conditions are imposed to a time $\eta_f$ during inflation when curvature perturbation are super horizon and hence their amplitude is frozen. 
Using Eqns. (\ref{eqn:tree}, \ref{eqn:Rtpf}) and (\ref{eqn:BispectrumR}), the expression for bispectrum of curvature perturbation can be written in terms of Mukhanov Sasaki variable $v_k(\eta)$ as
\begin{eqnarray}\label{eqn:BRexp}
 B_{\cal R}(k_1, k_2, k_3)\, &=&\, \frac{v_{k_1}(\eta_f)}{z(\eta_f)}\,\frac{v_{k_2}(\eta_f)}{z(\eta_f)}\,\frac{v_{k_3}(\eta_f)}{z(\eta_f)}
\,\int_{\eta_i}^{\eta_f} {\rm d}\eta'\, g(\eta')\, {\cal F}(v_{k_1}(\eta'),\, v_{k_2}(\eta')\, v_{k_3}(\eta'))\, \nonumber\\
&&+\, {\rm complex~ conjugate}.
\end{eqnarray}
In the above expression $g(\eta')$ is, in general, a function dependent on the background quantities and ${\cal F}(v_{k_1}(\eta'),\, v_{k_2}(\eta')\, v_{k_3}(\eta'))$ contains several terms which correspond to different terms in the third order Hamiltonian Eqn. (\ref{eqn:H3}). 
From figure \ref{fig:zppbyz}, it is clear that modes whose wavenumber is much larger than $k_{_{\rm LQC}}$ will not be affected by the bounce. These modes remain in Bunch-Davies vacuum at the onset of slow roll inflation. 
Hence, these modes will have a nearly scale invariant bispectrum with an amplitude of the order of slow roll parameters \cite{Maldacena:2002vr}. In this work, we approximate the shape of bispectrum of wavenumbers $k >> k_{_{\rm LQC}}$ by using the local template, see, for instance \cite{Komatsu:2001rj},
\begin{eqnarray}\label{eqn:Blocal}
 B_{\cal R}^{\,\rm local}(k_1, k_2, k_3)\, &=&\, - \, \, \frac{6}{5}\,(2\pi^2)^2\,{\mathfrak f}^{\rm \,local}_{_{\rm NL}}\,\biggl(\, \frac{{\cal \widetilde P_R}(k_1)}{k_1^3}\,\frac{{\cal \widetilde P_R}(k_2)}{k_2^3}\,
+\, \frac{{\cal \widetilde P_R}(k_2)}{k_2^3}\,\frac{{\cal \widetilde P_R}(k_3)}{k_3^3}\, +\, \frac{{\cal \widetilde P_R}(k_3)}{k_3^3}\,\frac{{\cal \widetilde P_R}(k_1)}{k_1^3} \biggr),\nonumber\\
\end{eqnarray}
where ${\cal \widetilde P_R}(k_1)\, =\, A_s (k/k_\star)^{n_s-1}$ and we fix ${\mathfrak f}^{\rm \,local}_{_{\rm NL}}\, =\, 10^{-2}$.
\par
Let us now try to understand the bispectrum of modes with wavenumbers $k$ comparable to or smaller than $k_{_{\rm LQC}}$. To estimate the bispectrum of these modes which are affected by the bounce, let us first focus on the integral involved. The integral is computed from a time $\eta_i$ to $\eta_f$. To isolate and understand the contribution due to the bounce, we will focus only on the contribution to this integral from a time interval around the bounce, {\it viz.~} $\eta \in [-\eta_0, \eta_0]$, where $\eta_0$ is some finite time. We will now approximate ${\cal F}(v_{k_1}(\eta'),\, v_{k_2}(\eta')\, v_{k_3}(\eta'))$ by just ${\cal F}\, =\, v_{k_1}^*(\eta')\,v_{k_2}^*(\eta')\,v_{k_3}^*(\eta')$. In writing this expression, we ignore any factors of wavenumbers that may arise due to spatial derivatives in the Hamiltonian. We also ignore any time derivative of the perturbations. We further approximate the mode $v_k(\eta')$  by ${\rm e}^{-i\,k\,\eta'}$. This approximation is exact for modes $k>>k_{_{\rm LQC}}$. Thus, the integral in Eqn. (\ref{eqn:BRexp}) becomes 
\begin{eqnarray}\label{eqn:integral} 
{\cal I}\, &=&\, \int_{-\eta_0}^{\eta_0}\, {\rm d}\eta' \,g(\eta') {\rm e}^{i\,(k_1\,+\,k_2\,+\,k_3)\eta'}  \nonumber\\
&=&\, \int_{-\infty}^{\infty}\, {\rm d}\eta' \,g(\eta') {\rm e}^{i\,(k_1\,+\,k_2\,+\,k_3)\eta'}\, W(|\eta'-\eta_0|).
\end{eqnarray}
In the last line of the above equation, we have changed the limits of the integral by using a window function that is one only if $|\eta' - \eta_0|<0$ or else zero. 
\par
In this form, one can use Cauchy's residue theorem to evaluate the integral. According to this theorem, integral evaluates to $2\,\pi\,i$ times the sum of residues. Thus, in order to evaluate the integral, we need to analyze the poles of the background function $g(\eta')$. If $\eta_p$ is the pole of $g(\eta')$, then the integral will be proportional to ${\rm e}^{i\,(k_1\,+\,k_2\,+\,k_3)\,\eta_p}$. From this expression, we can see that real part of the pole would lead to an oscillatory behaviour and the imaginary part will lead to an exponential behaviour. Looking at the numerical results plotted in figure \ref{fig:fNL}, we can see that the bispectrum is indeed an exponentially decreasing oscillatory function of wavenumbers. We can proceed a step further and try to analyse the poles of $g(\eta')$ and trace the origin of the imaginary pole. For this, note that of the various background functions contained in $g(\eta')$,  $a(\eta')$ has a minimum at the bounce. If we Taylor expand $a(\eta')$ around the bounce, then $a(\eta')\, =\, a(\eta_b)\, +\, a''(\eta_b)\,\eta^{'2}$. Since $a(\eta')$ has a minimum at bounce, $a''(\eta_b) > 0$. 
This implies that $g(\eta')$ will have an imaginary pole at $\eta_p\, =\, \sqrt{-a(\eta_b)/a''(\eta_b)}$. Thus, the imaginary poles arise from factors of $1/a(\eta')$ in the integrand. 
From the expression for scale factor near the bounce in LQC, see, for instance, \cite{Bolliet:2015bka},
\begin{equation}
 a(t)\, =\, a_B\left( 1\, +\, 3\,\kappa\,\rho_{sup}\,t^2\right)^{1/6},
\end{equation}
we obtain the pole to be at $t_p\,=\, i/\sqrt{(3\,\kappa\,\rho_{sup})}$. 
In terms of conformal time,
\begin{equation}
 \eta_p\, = \, \int_0^{t_p}\,\frac{{\rm d}t'}{a(t')}\, =\, \frac{t_p}{a_B}\, {}_2F_1\,\left[\frac{1}{6}, \frac{1}{2}, \frac{3}{2}, -3\,\kappa\,\rho_{sup}\,t_p^2\right]\, = \, i\,\frac{0.647}{k_{_{\rm LQC}}}.
\end{equation}
\par
Thus, the scale dependence of the integral due to the imaginary pole can be approximated to be  
 ${\rm e}^{i\,(k_1\,+\,k_2\,+\,k_3)\,\eta_p}\,=\,{\rm e}^{-0.647\,\frac{k_1\,+\,k_2\,+\,k_3}{k_{_{\rm LQC}}}}$.
The real poles could lead to an oscillatory behaviour. We model  the oscillatory behaviour of the bispectrum (obtained after adding the complex conjugate) as $\frac{\sin(k_1\,+\,k_2\,+\,k_3)}{k_I}$.
\par
Finally, the bispectrum also contain a factor of $\frac{v_{k_1}(\eta_f)}{z(\eta_f)}\,\frac{v_{k_2}(\eta_f)}{z(\eta_f)}\,\frac{v_{k_3}(\eta_f)}{z(\eta_f)}$.
Since, $v_k(\eta_f)/z(\eta_f)\, =\, {\cal R}_k(\eta_f)$, we substitute 
\begin{equation}
 \frac{v_{k}(\eta_f)}{z(\eta_f)}\,\approx\, \sqrt{\frac{{\cal P_R}(k)}{k^3}}.
\end{equation}
Putting everything together, the bispectrum of curvature perturbations in the regime $k \lesssim k_{_{\rm LQC}}$ can be approximated using the following template,
\begin{eqnarray}\label{eqn:Bbounce}
 B_{\cal R}^{\,\rm bounce}(k_1, k_2, k_3)\, &=&\,-\,\frac{6}{5}\,(2\pi^2)^2\,{\mathfrak f}^{\rm \,bounce}_{_{\rm NL}}\, \biggl( \frac{{\cal P_R}(k_1)}{k_1^3}\,\frac{{\cal P_R}(k_1)}{k_1^3}\,\frac{{\cal P_R}(k_1)}{k_1^3}\biggr)^{1/2}
 \,{\rm e}^{-0.647\,\frac{k_1\,+\,k_2\,+\,k_3}{k_{_{\rm LQC}}}}\nonumber\\
 &&\times\, \,\sin\left( \frac{k_1\,+\,k_2\,+\,k_3}{k_I} \right) .
\end{eqnarray}
In the above expression, we fix ${\mathfrak f}^{\rm \,bounce}_{_{\rm NL}}\, =\, 1\,{\rm M_{Pl}}^{-3/2}$. The factor of $-\frac{6}{5}(2\pi^2)^2$ has been included to make the expression consistent with the conventional definition of $\fnl$ (see, for instance, \cite{Komatsu:2001rj}).
\par 
Thus, the complete template for bispectrum of curvature perturbation generated in LQC is 
\begin{equation}\label{eqn:Btemplate}
 B_{\cal R}^{\,\rm LQC}(k_1, k_2, k_3)\,=\, B_{\cal R}^{\,\rm local}(k_1, k_2, k_3)\,+\, B_{\cal R}^{\,\rm bounce}(k_1, k_2, k_3)\,.
\end{equation}
In writing the above expression, we have assumed that the oscillations in the bispectrum are about the bispectrum due to the local template. 
The non-Gaussianity function corresponding to the above template for bispectrum can be computed from Eqn. (\ref{eqn:fNL}). 
A comparison of the dimensionless non-Gaussianity function corresponding to the above template with the numerical results in three different configuration of wavenumbers is given in figure \ref{fig:fNL}. We find that this simple template of bispectrum qualitatively captures the essential features of the primordial non-Gaussianity generated in LQC. 
Hence we will use the templates in Eqns. (\ref{eqn:PStemplate}, \ref{eqn:Btemplate}) to compute the reduced bispectrum generated in LQC.


%
\section{CMB bispectra\label{sec:cmb}}
Primordial perturbations leave their imprints in the CMB radiation as temperature fluctuations and as electric and magnetic polarisations, often referred to as E and B modes respectively (see, for instance, \cite{durrer_2020, Dodelson:2003ft, Weinberg_2008}). The temperature fluctuations and E modes are produced from primordial scalar perturbations, whereas B modes are not. 
Since we are interested in understanding the imprints of  scalar bispectrum, we will focus on the bispectra of temperature fluctuations and electric polarisation and their three-point cross-correlations. In this section, we will discuss the essential aspects of computing these bispectra.
\par 
Since CMB is observed on a sphere, namely the surface of last scattering, it is convenient to decompose it in terms of spherical harmonics,
\begin{equation}
	X(\hat n)\, =\, \sum_{\ell, m} a^{\text{\tiny X}}_{_{\ell m}}\, Y_{\ell m}(\hat n)
\end{equation}
where $X$ could be either fluctuation in temperature defined as $(T(\hat n) - \bar{T})/\bar{T}$, where $\bar T$ is the mean temperature of the CMB, or electric polarisation $E(\hat n)$. 
The multipole $a_{_{\ell m}}^{\text{\tiny X}}$ corresponding to anisotropies in the temperature and electric polarisation is related to the curvature perturbation through the relation
\begin{equation}
	a_{_{\ell m}}^{\text{\tiny X}} = 4\pi\,(-i)^{\ell}\, \int \frac{{\rm d}^{3}k}{(2\pi)^{3}}\, {\cal R}_k\,\Delta_{_\ell}^{\text{\tiny X}}(k)\,Y_{_{\ell m}}(k).
\end{equation}
In the above, $\Delta^{\text{\tiny X}}_{_\ell}$ is the transfer function which captures the physics post horizon exit of perturbations towards the end of inflation. 
We are interested in calculating the three-point function of these multipoles of the form $\langle a_{_{\ell_1 m_1}}^{\text{\tiny X}}\,a_{_{\ell_2 m_2}}^{\text{\tiny Y}}\,a_{_{\ell_3 m_3}}^{\text{\tiny Z}}\rangle$, where $X$, $Y$ and $Z$ can be either temperature fluctuations or E mode polarisation and where the average is over different realisations of the Universe. 
\par 
The three-point function of multipole coefficients can be expressed in terms of three-point functions of primordial perturbations as \cite{Komatsu:2001rj, Fergusson:2006pr, 2010AdAst2010E..73L, 2012JCAP...12..032F, durrer_2020}
\begin{eqnarray}
	\langle a_{_{{\ell_1} {m_1}}}^{\text{\tiny X}}\, a_{_{{\ell_2} {m_2}}}^{\text{\tiny Y}}\, a_{_{{\ell_3} {m_3}}}^{\text{\tiny Z}}\,\rangle\,& =&\, \left( 4\pi\right)^3\, \left(- i\right)^{\ell_1+\ell_2+\ell_3}\, \int \frac{\d^3 k_1}{(2\pi)^3} \int \frac{\d^3 k_2}{(2\pi)^3} \int \frac{\d^3 k_3}{(2\pi)^3} \Delta_{_{\ell_1}}^{\text{\tiny X}} \Delta_{_{\ell_2}}^{\text{\tiny Y}} \Delta_{_{\ell_3}}^{\text{\tiny Z}}\, \nonumber\\
	&\times&\langle {\cal R}_{k_1} {\cal R}_{k_2} {\cal R}_{k_3}\rangle\, Y_{_{\ell_1 m_1}} (\hat k_1)\, Y_{_{\ell_2 m_2}} (\hat k_2)\, Y_{_{\ell_3 m_3}} (\hat k_3).
\end{eqnarray}
Using Eqn. (\ref{eqn:BispectrumR}) and expressing the Dirac-Delta function in its exponential form, we obtain 
\begin{eqnarray}
		\langle a_{_{{\ell_1} {m_1}}}^{\text{\tiny X}}\, a_{_{{\ell_2} {m_2}}}^{\text{\tiny Y}}\, a_{_{{\ell_3} {m_3}}}^{\text{\tiny Z}}\,\rangle\,& =&\, \left( 4\pi\right)^3\, \left(- i\right)^{\ell_1+\ell_2+\ell_3}\, \int \frac{\d^3 k_1}{(2\pi)^3} \int \frac{\d^3 k_2}{(2\pi)^3} \int \frac{\d^3 k_3}{(2\pi)^3} \Delta_{_{\ell_1}}^{\text{\tiny X}} \Delta_{_{\ell_2}}^{\text{\tiny Y}} \Delta_{_{\ell_3}}^{\text{\tiny Z}}\, \nonumber\\
		&\times& \int \d^3 x\, e^{i(\vec k_1\,+\,\vec k_2\,+\,\vec k_3).\vec x}\, B_{\mathcal{R}}(k_1,k_2,k_3)\,Y_{_{\ell_1 m_1}} (\hat k_1)\, Y_{_{\ell_2 m_2}} (\hat k_2)\, Y_{_{\ell_3 m_3}} (\hat k_3).\nonumber\\
\end{eqnarray}
Up on using plane wave expansion,
\begin{equation}
	e^{i\, \vec k\cdot \vec x}\, =\, \sum_{\ell = 0}^{\infty}\, \sum_{m = -\ell}^{\ell}\, i^\ell\, j_\ell (k\,x)\, Y_{\ell m}(\hat x)\,Y^*_{\ell m}(\hat k),
\end{equation}
and the orthonormal property of spherical harmonics, we obtain
\begin{equation}
	\langle a_{_{{\ell_1} {m_1}}}^{\text{\tiny X}}\, a_{_{{\ell_2} {m_2}}}^{\text{\tiny Y}}\, a_{_{{\ell_3} {m_3}}}^{\text{\tiny Z}}\,\rangle\, =\, b_{\,\ell_1\,\ell_2\,\ell_3}^{\text{\tiny XYZ}}\, {\cal G}^{{m_1\,m_2\,m_3}}_{\ell_1\, \ell_2\, \ell_3},
\end{equation}
where all the dependence on $m$ indices are captured in the Gaunt integral
\begin{equation}
	 {\cal G}^{{m_1\,m_2\,m_3}}_{\ell_1\, \ell_2\, \ell_3}\, =\, \int \d \hat x\, Y_{\ell_1 m_1}(\hat x)\, Y_{\ell_2 m_2}(\hat x)\, Y_{\ell_3 m_3}(\hat x).
\end{equation}
The quantity $b_{\,\ell_1\,\ell_2\,\ell_3}^{\text{\tiny XYZ}}$ is called the reduced bispectrum and is given by
\begin{eqnarray}\label{eqn:bl1l2l3}
	b_{\,\ell_1\,\ell_2\,\ell_3}^{\text{\tiny XYZ}} \, &=&\, \left(\frac{2}{\pi}\right)^3\, \int x^2 \d x\, \int \d k_1\, \int \d k_2\, \int \d k_3\, \left( k_1\,k_2\,k_3\right)^2\, B_{\mathcal{R}}(k_1,k_2,k_3)\,\nonumber\\
	&\times&\, \Delta_{_{\ell_1}}^{\text{\tiny X}} \Delta_{_{\ell_2}}^{\text{\tiny Y}} \Delta_{_{\ell_3}}^{\text{\tiny Z}}\, j_{\ell_1} (k_1\,x)\, j_{\ell_2} (k_2\,x)\, j_{\ell_3} (k_3\,x).
\end{eqnarray}
The presence of Gaunt integral implies that the reduced bispectra is non-zero only when the multipoles satisfies the triangle inequality $|\ell_{1}-\ell_{2}| \leq \ell_{3}  \leq |\ell_{1}+\ell_{2}|$ and when $\ell_1\, +\, \ell_2\, +\, \ell_3$ is even. 
For isotropic theories, it suffices to work with the reduced bispectrum.  
\section{Reduced bispectra from loop quantum cosmology}\label{sec:lqcbispectra}
We will now compute the reduced bispectrum generated in LQC. The reduced bispectrum corresponding to 
the primordial bispectrum Eqn. (\ref{eqn:Btemplate}) can be computed using Eqn. (\ref{eqn:bl1l2l3}). 
However, this calculation involves four integrals, three  over wavenumbers and one over $x$ variable, which is computationally expensive. 
This calculation can however be simplified, essentially just to two integrals, if we use the separable property of the above primordial bispectrum. The total primordial bispectrum is not separable, however, the contribution due to the bounce and that due to the scale invariant local template are separable. Hence, the reduced bispectrum in LQC can be expressed as  
\begin{eqnarray} \label{eqn:bl1l2l3LQC}
	  b_{{\ell_{1}}{\ell_{2}}{\ell_{3}}}^{\text{\tiny{XYZ}}}&=&\, b_{{\ell_{1}}{\ell_{2}}{\ell_{3}}}^{{\rm bounce}}\, +\, b_{{\ell_{1}}{\ell_{2}}{\ell_{3}}}^{{\rm local}},
\end{eqnarray}
where
\begin{eqnarray}\label{eqn:sepbounce}
	b_{{\ell_{1}}{\ell_{2}}{\ell_{3}}}^{{\rm bounce}}\,&=&\,
	-\bigg(\frac{2}{\pi}\bigg)^{3}\frac{6}{5}\,(2\pi^{2})^{2}\,{\mathfrak f}^{\rm \,bounce}_{_{\rm NL}}\,\int_{0}^{\infty} d x\, x^2\, \biggl[A^{^X}_{\ell_{1}}(x)\,B^{^Y}_{\ell_{2}}(x)\,B^{^Z}_{\ell_{3}}(x)\nonumber\\
	&+&  
	B^{^X}_{\ell_{1}}(x)\,A^{^Y}_{\ell_{2}}(x) 
	B^{^Z}_{\ell_{3}}(x)\,+\, B^{^X}_{\ell_{1}}(x)\,B^{^Y}_{\ell_{2}}(x)\,A^{^Z}_{\ell_{3}}(x)\,- A^{^X}_{\ell_{1}}(x)\,A^{^Y}_{\ell_{2}}(x)\,A^{^Z}_{\ell_{3}}(x)\biggr],\nonumber\\ 
	\end{eqnarray}
and
\begin{eqnarray}\label{eqn:seplocal}
	b_{{\ell_{1}}{\ell_{2}}{\ell_{3}}}^{{\rm local}} &=&\,
	-\bigg(\dfrac{2}{\pi}\bigg)^{3}\, \frac{6}{5}\,(2\pi^{2})^{2}\,{\mathfrak f}^{\rm \,local}_{_{\rm NL}}\,\int_{0}^{\infty} \d x\, x^2\,\biggl[E^{^X}_{\ell_{1}}(x)\,E^{^Y}_{\ell_{2}}(x)\,G^{^Z}_{\ell_{3}}(x) \nonumber
	+ G^{^X}_{\ell_{1}}(x)\,E^{^Y}_{\ell_{2}}(x)\,E^{^Z}_{\ell_{3}}(x) \nonumber\\ 
	&+& E^{^X}_{\ell_{1}}(x)\,G^{^Y}_{\ell_{2}}(x)\,E^{^Z}_{\ell_{3}}(x)\biggl].
    \end{eqnarray}
In the above expressions, the functions $A^{^X}_\ell(x)$, $B^{^X}_\ell(x)$, $E^{^X}_\ell(x)$ and $G^{^X}_\ell(x)$ are 
\begin{subequations}\label{eqn:ABEG}
\begin{eqnarray}
	A^{^X}_{\ell}(x) &=&\int_{0}^{\infty} dk\,\Delta_{\ell}^{\text{\tiny X}}(k)\,j_{\ell}(kx)\,\sqrt{({\cal P}_{\cal R}(k) {k}}\,\mathrm{e}^{-0.647\frac{k}{k_{_{\rm LQC}}}}\,\sin({\frac{k}{k_{_I}}}),\\
	B^{^X}_{\ell}(x) &=& \int_{0}^{\infty} dk\,\Delta_{\ell}^{\text{\tiny X}}(k)\,j_\ell(kx)\,\sqrt{{\cal P}_{\cal R}(k){k}}\,\mathrm{e}^{-0.647\frac{k}{k_{_{\rm LQC}}}}\,\cos({\frac{k}{k_{_I}}}),\\
	E^{^X}_{\ell}(x) &=& \int_{0}^{\infty} dk\,\Delta_{\ell}^{\text{\tiny X}}(k)\,j_\ell(kx)\,k^{-1}\, {\cal \widetilde P_R}(k_1),\\
	G^{^X}_{\ell}(x) &=& \int_{0}^{\infty} dk\,\Delta_{\ell}^{\text{\tiny X}}(k)\,j_\ell(kx)\,k^{2}.
\end{eqnarray}
\end{subequations}
\par 
Note that, each of the functions $A^{^X}_\ell (x)$, $B^{^X}_\ell (x)$, $E^{^X}_\ell (x)$ and $G^{^X}_\ell (x)$ involve an integral over the wavenumber. The reduced bispectrum can now be calculated by evaluating these functions for all the required values of multipoles and then finally performing the integrals Eqns. (\ref{eqn:sepbounce}) and (\ref{eqn:seplocal}).

\section{Numerical procedure and results}\label{sec:numerics}
\begin{figure}
 \begin{tabular}{cc}
  \includegraphics[width=0.45\textwidth]{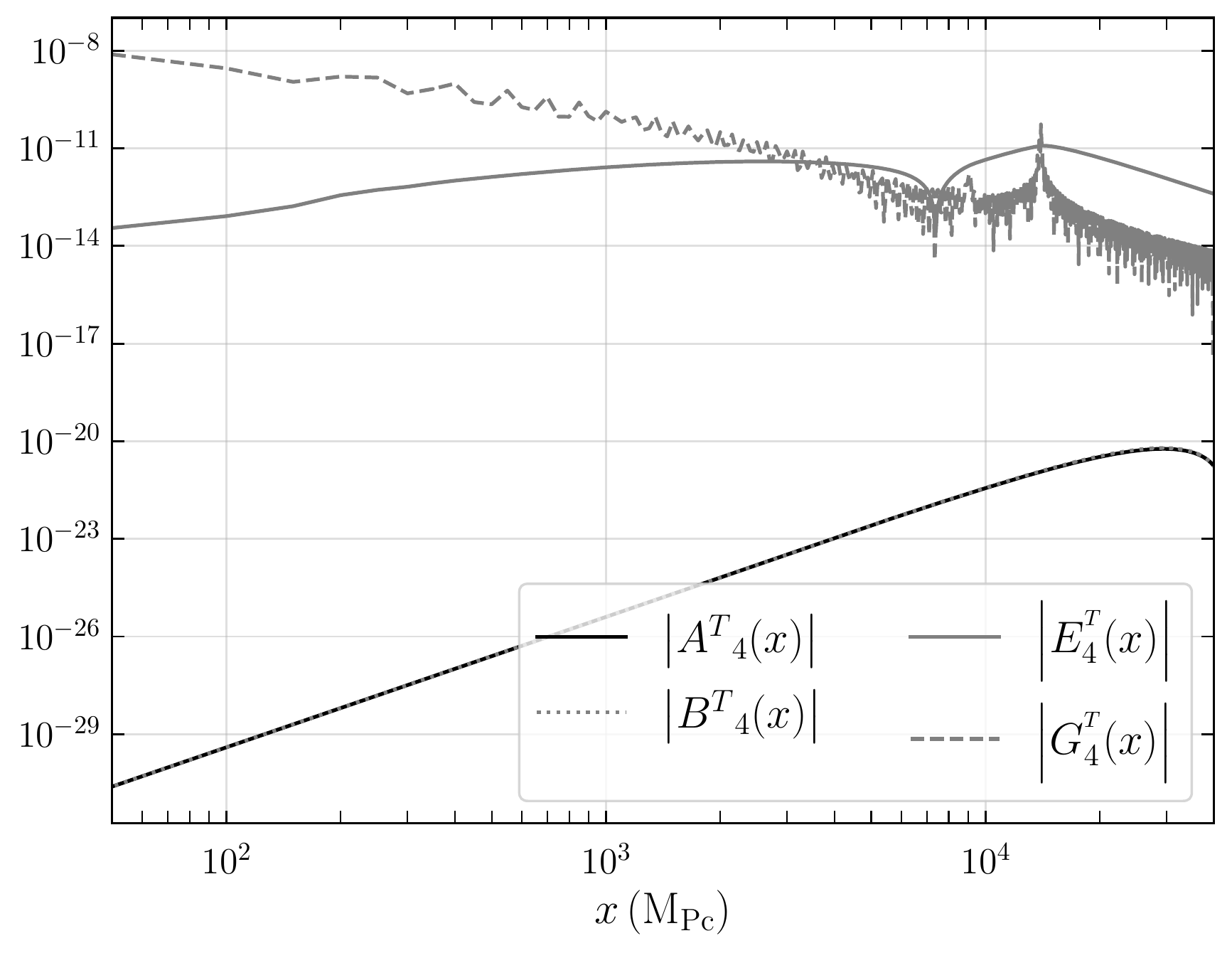}&
  \includegraphics[width=0.45\textwidth]{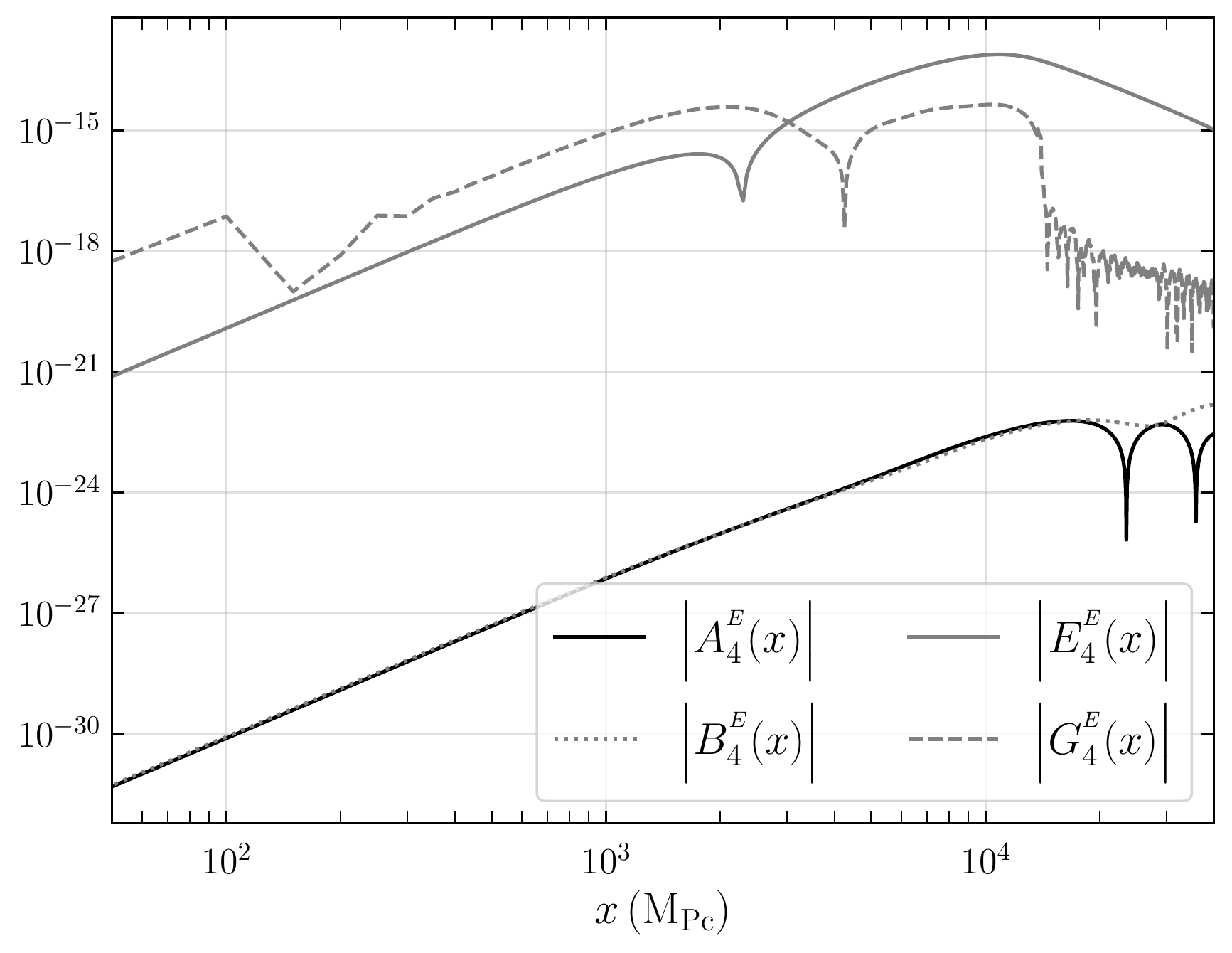}\\
   \includegraphics[width=0.45\textwidth]{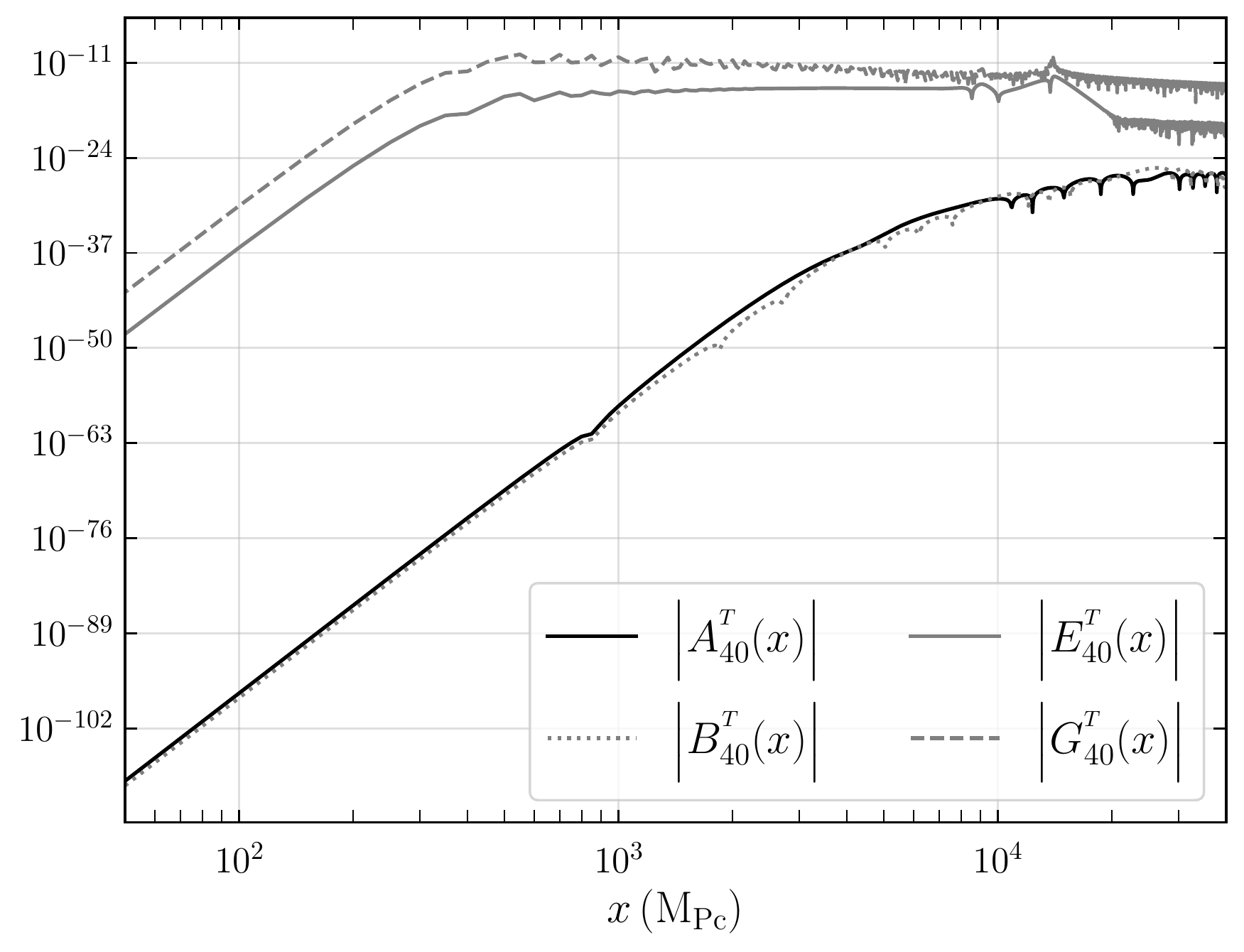}&
  \includegraphics[width=0.45\textwidth]{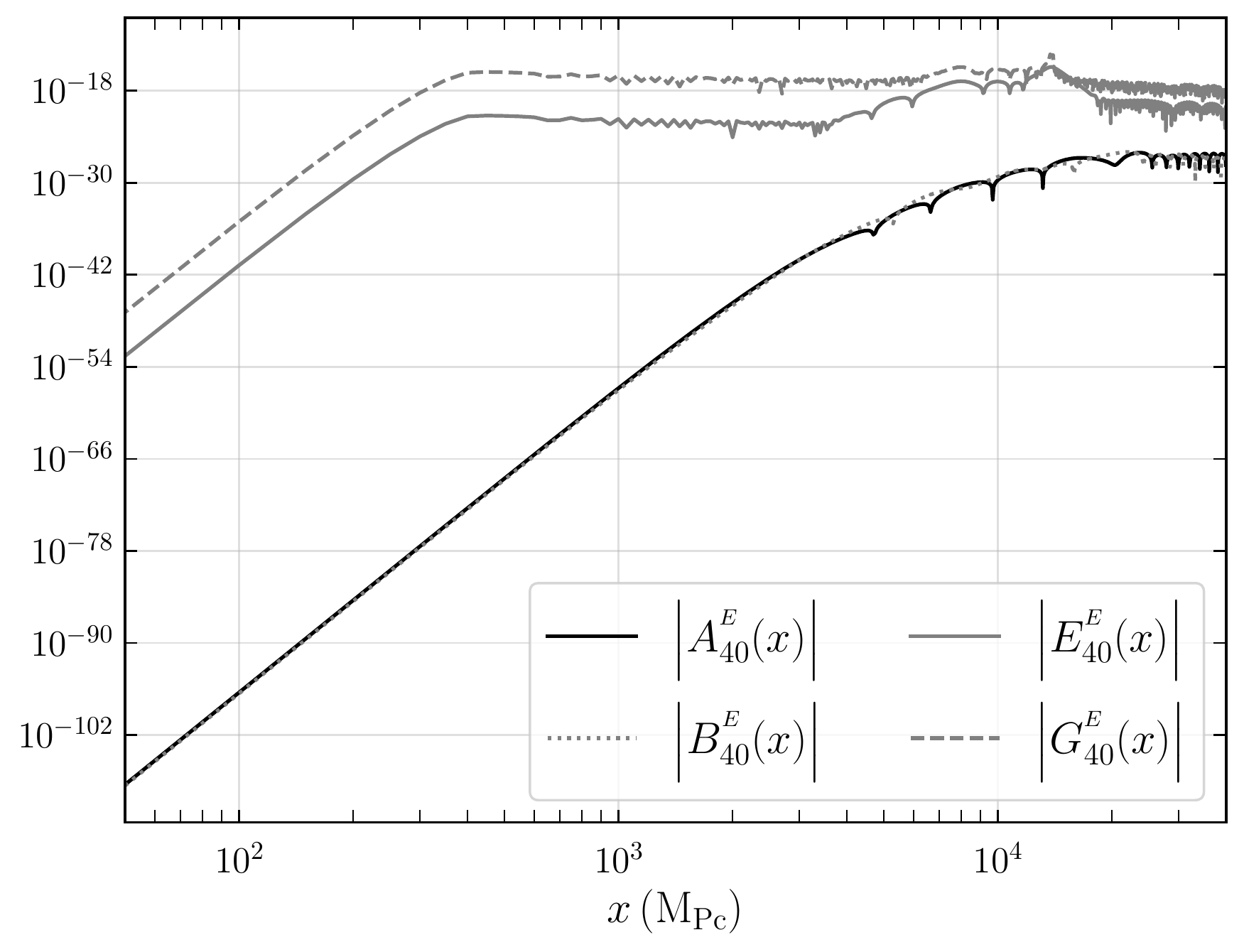}\\
 \end{tabular}
\caption{\label{fig:ABEG}The behaviour of functions in Eqn. (\ref{eqn:ABEG}) with $x$ for multipoles $\ell\, =\, 4$ and $40$. The contribution from the local part of the bispectrum {\it viz.~} $E^{^X}_\ell (x)$ and $G^{^X}_\ell (x)$ are clearly dominant compared to those arising from the bounce part {\it viz.~} $A^{^X}_\ell (x)$ and $B^{^X}_\ell (x)$. }
\end{figure}

We now discuss the numerical procedure we have followed for computing the reduced bispectrum generated in LQC. The first step in calculating reduced bispectrum is the evaluation of functions Eqns. (\ref{eqn:ABEG}). In order to compute these functions, we require the transfer functions $\Delta^{\text{\tiny X}}_\ell$, where $X$ can be either temperature fluctuations or electric polarisation. We use publicly available Boltzmann code \verb|class| \cite{2011JCAP...07..034B} to generate both the transfer functions. We perform the integral using Simpson's rule. We choose this method since the integrand is highly oscillatory and this gives better accuracy when we work with sufficiently small step size. Since the scale of oscillations occur at $k_I\,=\,10^{-7}\,{\rm Mpc}^{-1}$, we have worked with a step size of $\Delta k\, =\, 10^{-8}\,{\rm Mpc}^{-1}$. 
This leads to an accuracy of ${\mathcal O}(10^{-32})$. The behaviour of functions $A^{^X}_\ell (x)$, $B^{^X}_\ell  (x)$, $E^{^X}_\ell (x)$ and $G^{^X}_\ell (x)$ for multipoles $\ell\, =\, 4$ and $40$ are shown in figure \ref{fig:ABEG}. From the figure, it is clear that the functions $E^{^X}_\ell (x)$ and $G^{^X}_\ell (x)$ are dominant compared to $A^{^X}_\ell (x)$ and $B^{^X}_\ell (x)$. This is an indication of the fact that local part of the bispectrum is dominant compared to the oscillatory part due to the bounce. 
The next step in computing reduced bispectrum is the evaluation of Eqns. (\ref{eqn:sepbounce}, \ref{eqn:seplocal}). We perform these integrals over $x$ with a step size of $50$ in the range $x\,\in\, [0,\,40000]$.
We have made the calculations faster by using vectorization available in \verb|NumPy| and by parallelizing the computation wherever possible. 
\par
\begin{figure}
	\begin{tabular}{cc}
		\includegraphics[width=0.5\textwidth]{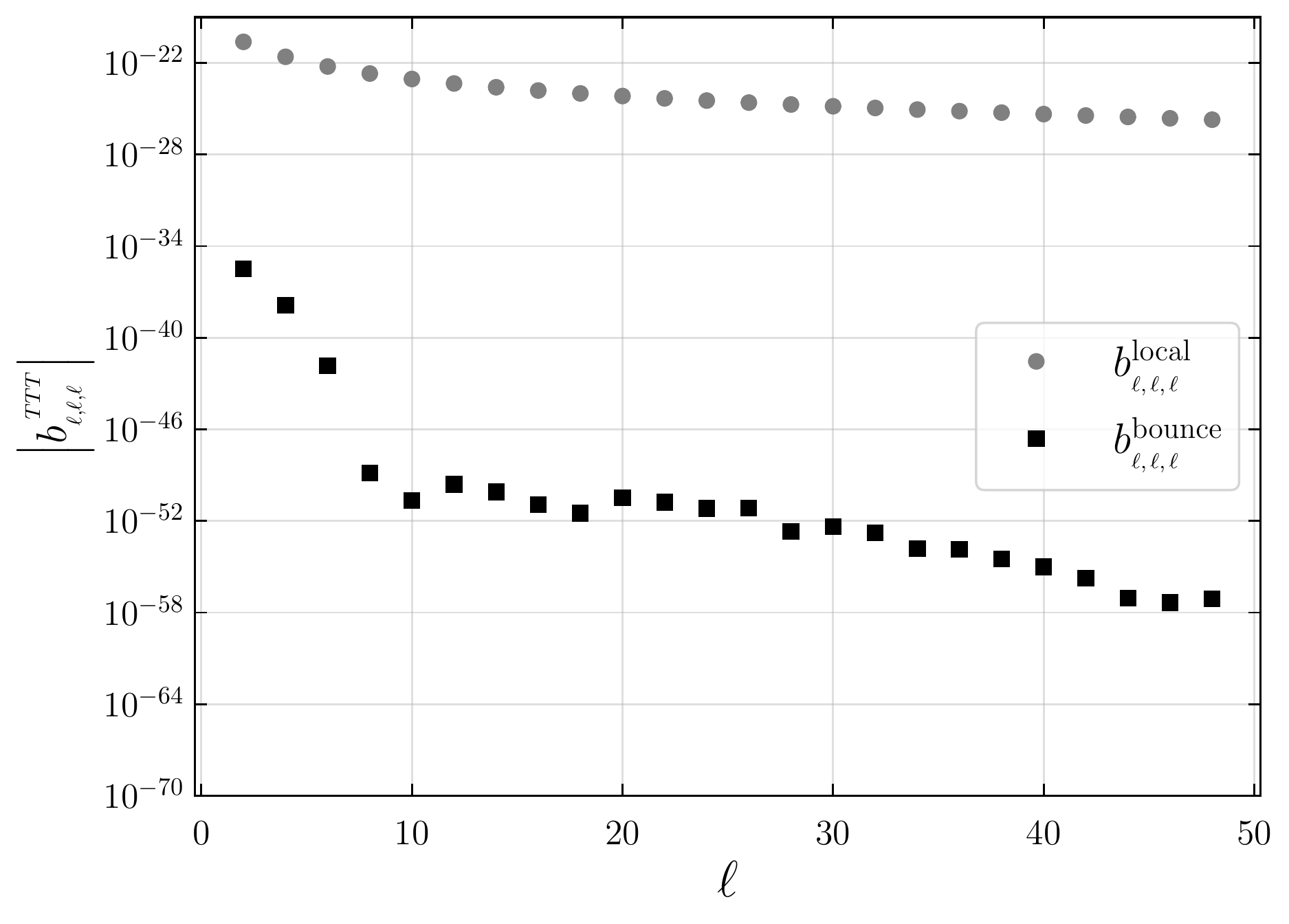}&
		\includegraphics[width=0.5\textwidth]{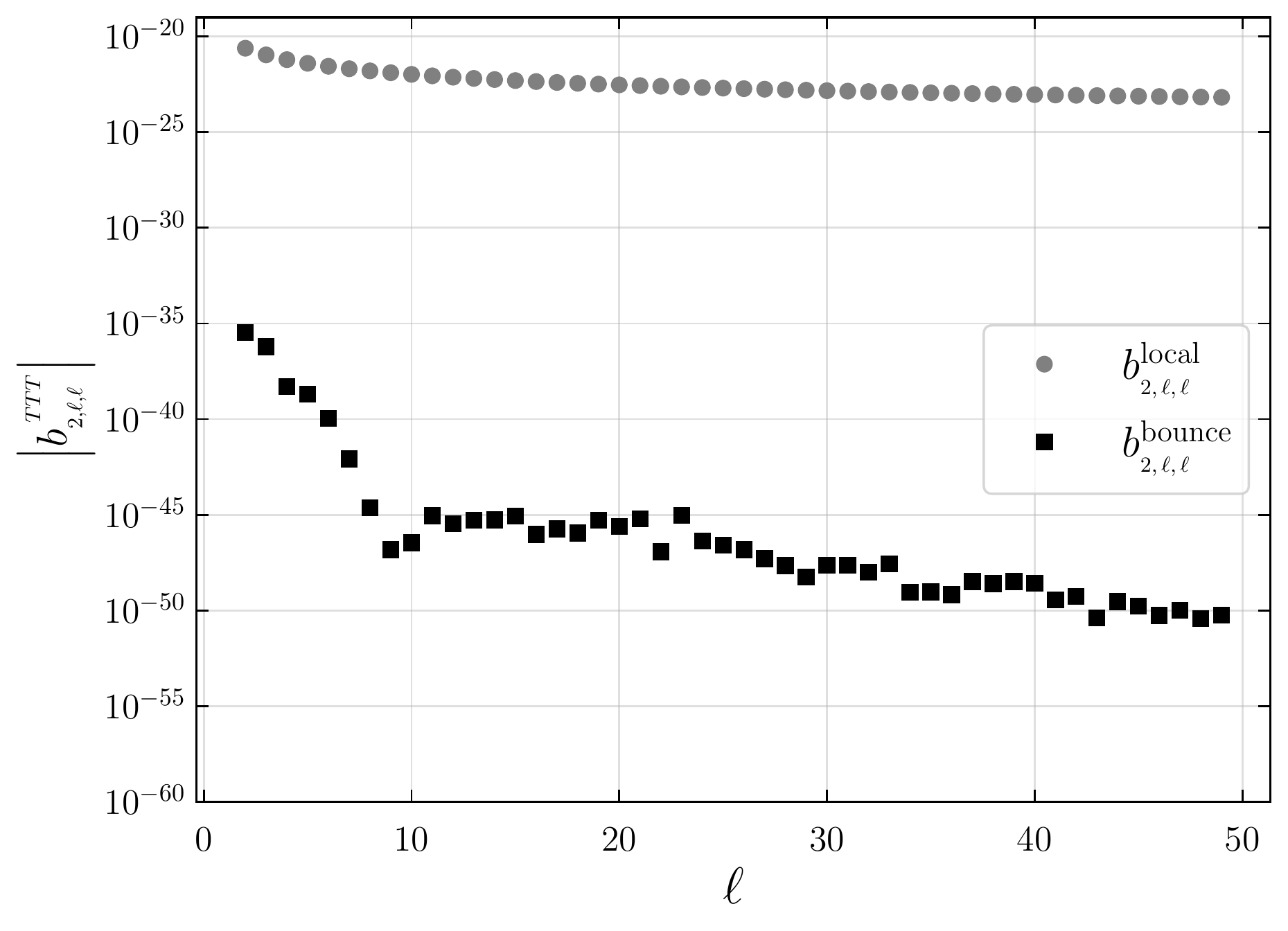}
	\end{tabular}
\caption{\label{fig:TTT} The reduced bispectra $b^{\text{\tiny TTT}}_{_{\,\ell_1,\,\ell_2,\,\ell_3}}$ in two different configurations. The plots illustrate that $b_{{\ell_{1}}{\ell_{2}}{\ell_{3}}}^{{\rm local}}\, >>\, b_{{\ell_{1}}{\ell_{2}}{\ell_{3}}}^{{\rm bounce}}$.}
\end{figure}

\begin{figure}
	\begin{tabular}{cc}
		\includegraphics[width=0.5\textwidth]{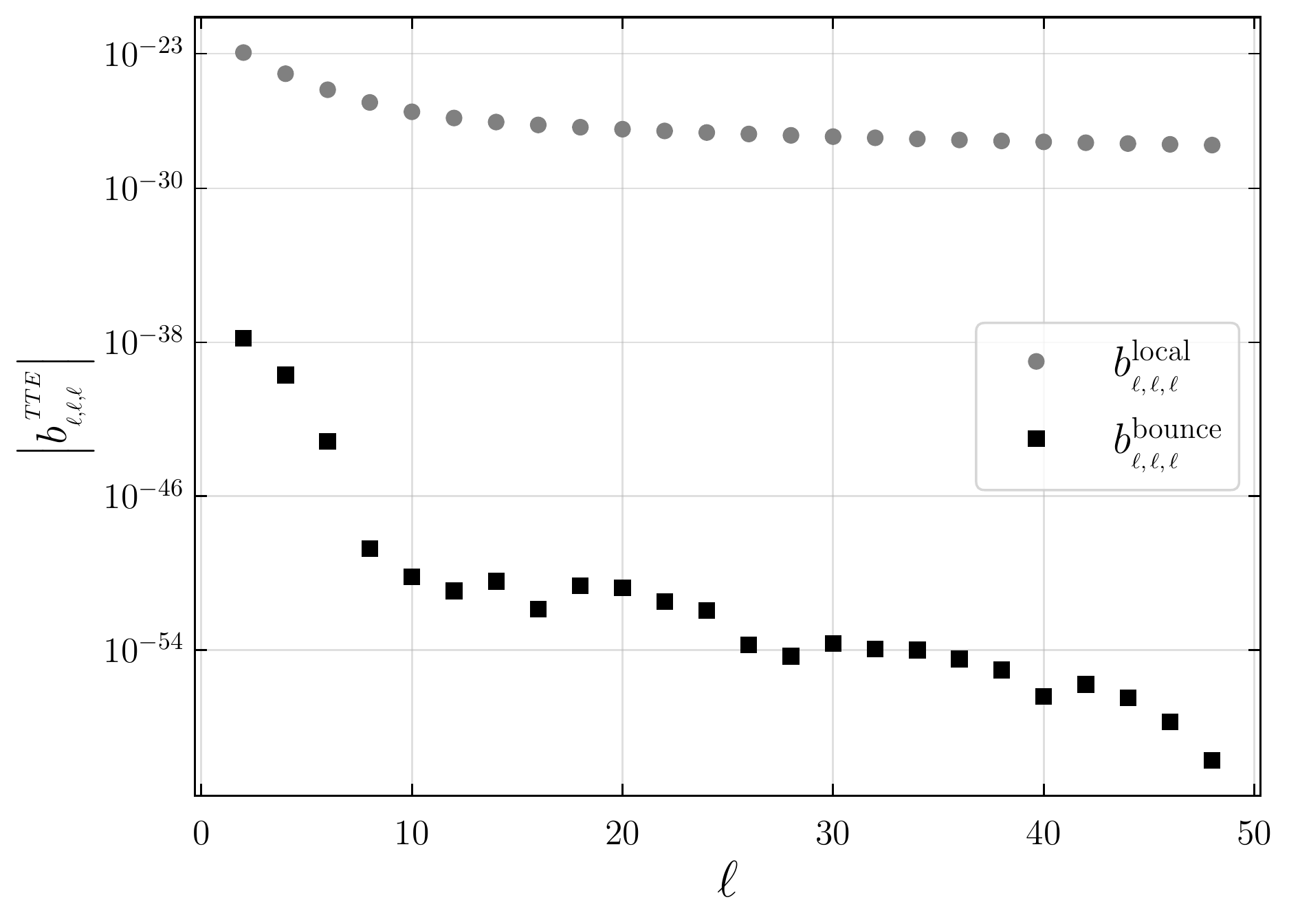}&
		\includegraphics[width=0.5\textwidth]{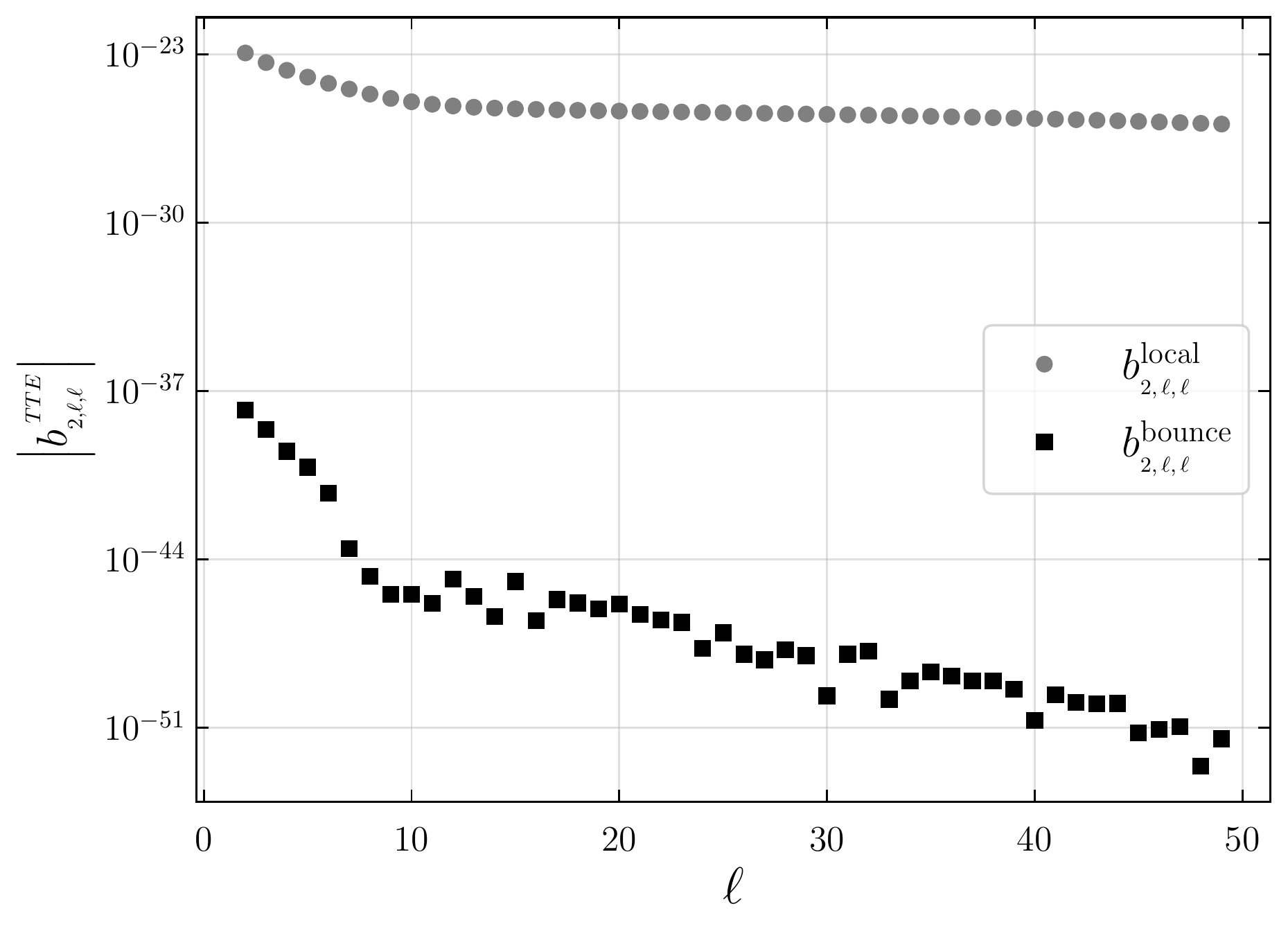}
	\end{tabular}
	\caption{\label{fig:TTE}The plots of reduced bispectra $b^{\text{\tiny TTE}}_{_{\,\ell_1,\,\ell_2,\,\ell_3}}$ in two different configurations. Note that $b_{{\ell_{1}}{\ell_{2}}{\ell_{3}}}^{{\rm local}}\, >>\, b_{{\ell_{1}}{\ell_{2}}{\ell_{3}}}^{{\rm bounce}}$.}
\end{figure}

\begin{figure}
	\begin{tabular}{cc}
		\includegraphics[width=0.5\textwidth]{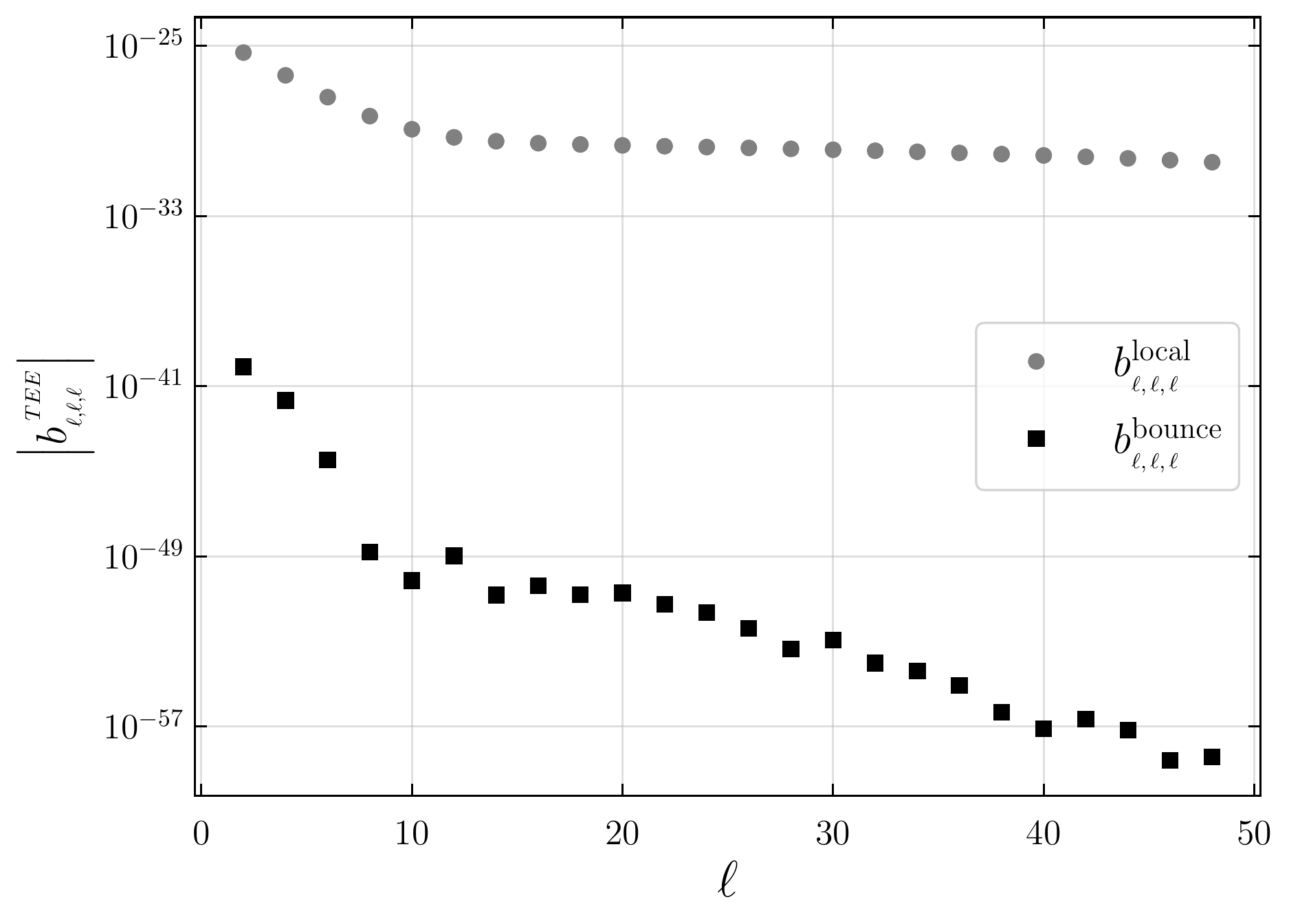}&
		\includegraphics[width=0.5\textwidth]{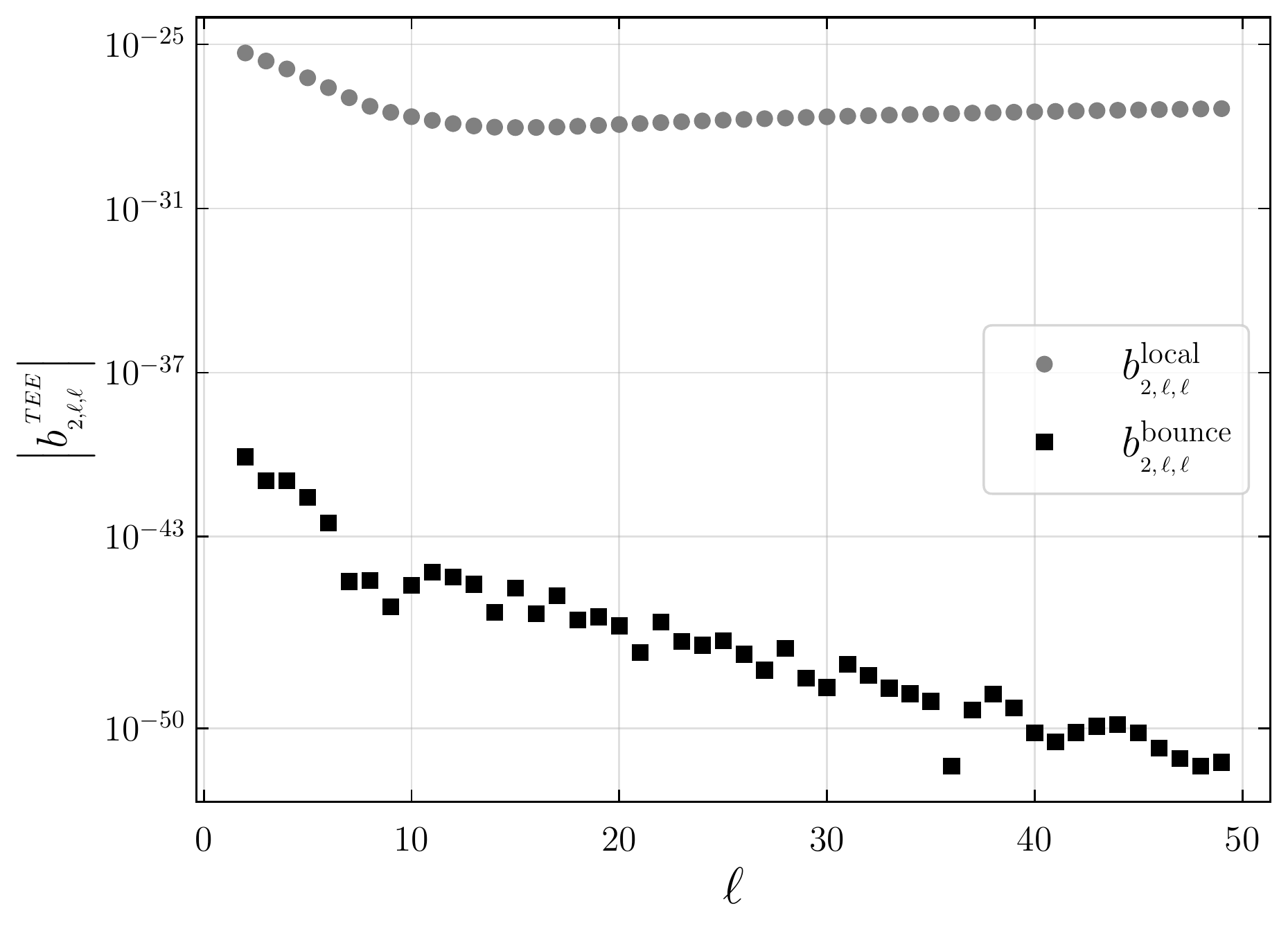}
	\end{tabular}
	\caption{\label{fig:TEE}The plots of reduced bispectra $b^{\text{\tiny TEE}}_{_{\,\ell_1,\,\ell_2,\,\ell_3}}$ in two different configurations. Clearly, the $b_{{\ell_{1}}{\ell_{2}}{\ell_{3}}}^{{\rm local}}$ is much larger than $b_{{\ell_{1}}{\ell_{2}}{\ell_{3}}}^{{\rm bounce}}$.}
\end{figure}

\begin{figure}
	\begin{tabular}{cc}
		\includegraphics[width=0.5\textwidth]{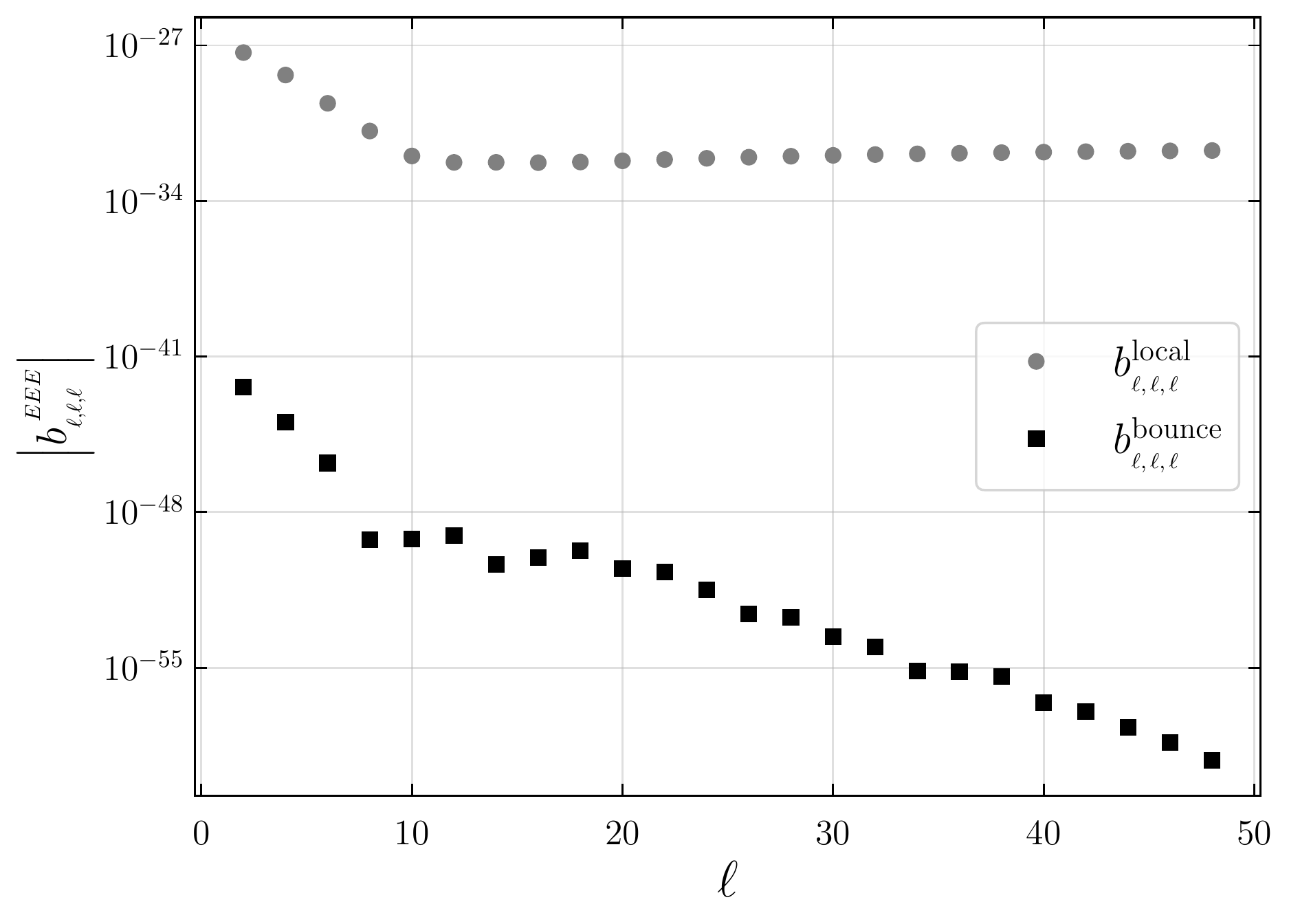}&
		\includegraphics[width=0.5\textwidth]{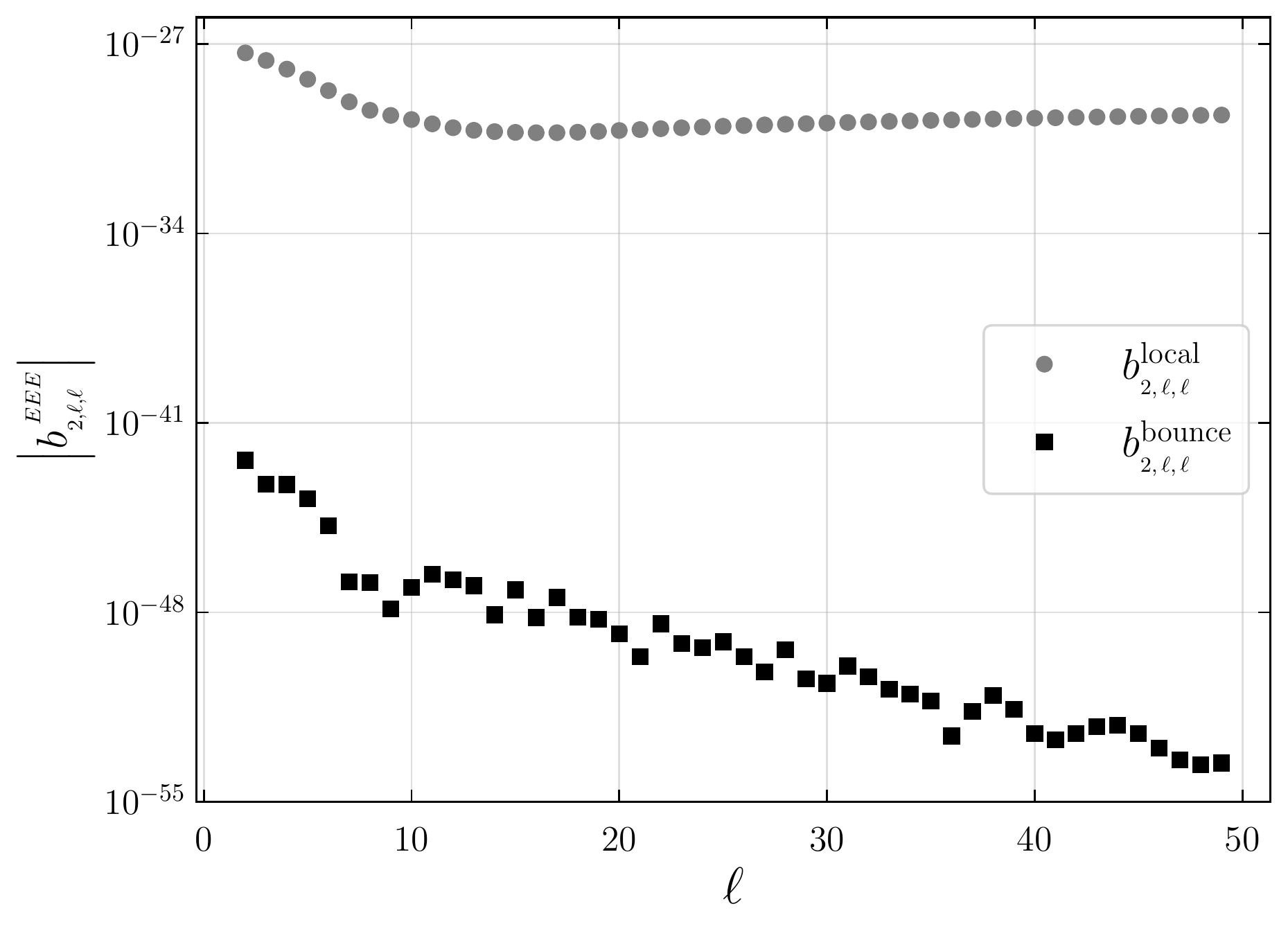}
	\end{tabular}
	\caption{\label{fig:EEE} The plots of reduced bispectra $b^{\text{\tiny EEE}}_{_{\,\ell_1,\,\ell_2,\,\ell_3}}$ in two different configurations. Note that, the reduced bispectrum is dominated by contribution from the local part of the template.}
\end{figure}

 
Reduced bispectra $b^{\text{\tiny TTT}}_{{\,\ell_1,\,\ell_2,\,\ell_3}}$, $b^{\text{\tiny TTE}}_{{\ell_1,\,\ell_2,\,\ell_3}}$, $b^{\text{\tiny TEE}}_{{\,\ell_1,\,\ell_2,\,\ell_3}}$ and $b^{\text{\tiny EEE}}_{{\,\ell_1,\,\ell_2,\,\ell_3}}$ generated in LQC are shown in figures \ref{fig:TTT}, \ref{fig:TTE}, \ref{fig:TEE} and \ref{fig:EEE} respectively. We have illustrated two different configurations of the bispectra. In these figures, we have separately plotted the contribution from the local and bounce parts of the bispectrum. The figures show that the contribution to the bispectrum from the oscillatory part of the template is negligible compared to that from the local part. This shows that the reduced bispectra generated in LQC will be similar to that produced in slow roll inflation and hence will be consistent with observations by Planck \cite{Akrami:2019izv}. 

\section{Summary and Discussion} \label{sec:discussion}
 
State of the art measurements of CMB by Planck has put strong constraints on primordial non-Gaussianity \cite{Akrami:2019izv}. Observations by Planck point towards a small primordial non-Gaussianity which is consistent with the one generated in slow roll models of inflation. In LQC, primordial perturbations  originate in an adiabatic vacuum before the bounce. These then evolve through the bounce, then through the inflationary epoch before their amplitude freezes upon horizon exit during inflation. The quantum bounce sets a scale $k_{_{\rm LQC}}$ in the problem. Modes which have wavenumbers comparable to or smaller than $k_{_{\rm LQC}}$ are excited during the bounce and modes with larger wavenumbers are not. This implies that modes with $k \lesssim k_{_{\rm LQC}}$ are in an excited and non-Gaussian state at the onset of inflation. This non-Gaussianity is then further enhanced as the modes exit the horizon during inflation. 
However, modes with longer wavenumbers, since they are not excited during the bounce, behave very similarly to modes in slow roll inflation. Hence, we have a situation where longer wavelength modes are strongly non-Gaussian whereas the shorter ones remain nearly Gaussian. In order to establish the viability of LQC as a model for pre-inflationary universe, it is important to answer whether LQC is compatible with the constraints on primordial non-Gaussianity set by Planck.
\par 
With this goal, we investigated the imprints of primordial non-Gaussianity in the bispectrum of temperature and electric polarisation and their cross-correlations generated in LQC. In particular, motivated by previous efforts, we proposed a template which captures the essential features of the primordial bispectrum generated in LQC Eqn. (\ref{eqn:Btemplate}). We then used the template to compute the $b^{\text{\tiny TTT}}_{_{\,\ell_1\,\ell_2\,\ell_3}}$, $b^{\text{\tiny TTE}}_{_{\,\ell_1\,\ell_2\,\ell_3}}$, $b^{\text{\tiny TEE}}_{_{\,\ell_1\,\ell_2\,\ell_3}}$ and $b^{\text{\tiny EEE}}_{_{\,\ell_1\,\ell_2\,\ell_3}}$ bispectra. To simplify the calculation, we used the separable property of the proposed template. 
The template for primordial bispectra in LQC presented in Eqn. (\ref{eqn:Btemplate}) consists of sum of two terms, namely
an exponentially decreasing oscillatory part $B^{\,\rm bounce}_{\cal R}(k_1,\, k_2,\, k_3)$ and a nearly scale invariant part $B_{\cal R}^{\,\rm local}(k_1,\, k_2,\, k_3)$. Corresponding to these two terms, the reduced bispectra given by Eqn. (\ref{eqn:bl1l2l3LQC}) consists of two terms, namely, $b_{{\ell_{1}}{\ell_{2}}{\ell_{3}}}^{{\rm bounce}}$ and $b_{{\ell_{1}}{\ell_{2}}{\ell_{3}}}^{{\rm local}}$. 
We find that $b_{{\ell_{1}}{\ell_{2}}{\ell_{3}}}^{{\rm bounce}}$ is negligible compared to $b_{{\ell_{1}}{\ell_{2}}{\ell_{3}}}^{{\rm local}}$. 
Thus, the reduced bispectrum generated in LQC will be dominated by the local part of the template, {\it viz.} $B_{\cal R}^{\,\rm local}(k_1,\, k_2,\, k_3)$ with an amplitude of the order of $\fnl \approx 10^{-2}$. 
This in turn implies that the reduced bispectrum generated in LQC is similar to that generated in slow roll inflation. Hence, we conclude that the primordial non-Gaussianity generated in LQC is compatible with the constraints from Planck. This is the central result of this paper. 
\par
The primordial perturbations generated in LQC is non-Gaussian in nature, yet our computation illustrates that the reduced bispectra of temperature and electric polarisation are similar to that of slow roll. This seemingly contradicting finding is because of the highly oscillatory nature of the primordial bispectrum. The reduced bispectra involves integrals over wavenumbers which average over these oscillations. Our result could be compared with those of \cite{Delgado:2021mxu, vanTent:2022vgy} where they had worked with a non-oscillatory template. While they found that, in the absence of oscillations, the contribution from the bounce is significant enough to be observed by Planck, we find that presence of oscillations in the primordial bispectra dilutes any imprints of non-Gaussianity on the reduced bispectra. Moreover, our result could also be compared to a purely oscillatory template, see, for instance, \cite{Chen:2006xjb, Munchmeyer:2014nqa}. The oscillation scale considered in such models is ${\cal O}\left( 10^{-3} - 10^{-2}\right)$ which leads to a multipole periodicity of roughly $14 - 140$. However, the scale of oscillations that we consider is $k_{\rm I}\, =\, 10^{-7}\,{\rm Mpc}^{-1}$. This corresponds to a multipole periodicity much smaller than one.  We believe this is the reason why the reduced bispectrum corresponding to the highly oscillatory part of our template is negligible. 
Our result, thus highlights the fact that a small reduced bispectra need not necessarily imply the absence of primordial non-Gaussianity. Hence, it would also be interesting to look for any other measurable imprints of such oscillatory and scale dependent primordial non-Gaussianity.
\par
Our findings are relevant for constraints on the amount of pre-inflationary expansion in LQC. In LQC, the amount of expansion before inflationary epoch is set by the value of scalar field at the bounce. The scalar field rolls up the potential after the bounce, comes to rest momentarily before it rolls down and settles in to the inflationary attractor. Hence, the value of the scalar field at the bounce determines the amount of expansion between the bounce and the onset of inflation. This epoch of expansion is relevant as it determines whether the scales that are sensitive to the effects of the bounce are visible today. 
If this epoch of pre-inflationary expansion is very large, then the imprints of the bounce will not be visible in the Universe today. However, if the pre-inflationary expansion is small, then the primordial power spectrum and bispectrum  will be scale dependent at observable scales. The power spectrum of temperature fluctuations fits extremely well to those due to a nearly scale invariant primordial power spectrum at multipoles $\ell > 30$ \cite{Aghanim:2018eyx, Aghanim:2019ame}. This imposes a lower limit to the amount of pre-inflationary expansion and hence a lower limit to the value of scalar field at the bounce \cite{Agullo:2015tca}. 
\par 
Compared to the primordial power spectrum, the primordial non-Gaussianity is more sensitive to the bounce, {\it i.e.~} $\fnl (k_1,\, k_2,\, k_3)$ is scale dependent at larger wavenumbers than the primordial power spectrum. This leads to a question whether imprints of primordial non-Gaussianity leads to a stronger lower limit to the pre-inflationary expansion. Our calculations, carried out in this work, answer this question in the negative. More specifically, since the reduced bispectra do not carry any imprints of the bounce, we find that it do not provide any constraints on the epoch of pre-inflationary expansion. 
\par
Finally, in this work, we have considered the imprints of the bounce in LQC for $\rho_{sup}\, =\, 0.41\,m_{_{\rm Pl}}^4$. This value corresponds to the lowest eigen value of the area operator obtained from calculations of black hole entropy in loop quantum gravity. We may take an alternate view and consider $\rho_{sup}$ to be a free parameter. As was shown in \cite{Agullo:2017eyh}, for a fixed value of scalar field at the bounce, a lower value of $\rho_{sup}$, will require lesser amount of pre-inflationary expansion from the bounce to the onset of inflation. This in turn would imply that the effects of the bounce are felt at shorter wavelengths for smaller values of $\rho_{sup}$. Moreover, the amplitude of the spectra will be lower for smaller values of $\rho_{sup}$. However, the qualitative scale-dependent oscillatory behaviour of the bispectrum will remain unchanged. Hence, we expect our conclusions to hold for other values of $\rho_{sup}$.
\section*{Acknowledgement}
We thank Ivan Agullo for his comments. This work was supported by Science and Engineering Research Board (SERB) through Start-up Research Grant SRG/2021/001769. We acknowledge the use of PU HPC facility of the National Supercomputing Mission for initial part of the project and thank Centre for Cyber Physical Systems, National Institute of Technology Karnataka, Surathkal for the financial support for the same.
\bibliographystyle{unsrt}
\bibliography{Refs}

\begin{thebibliography}{10}

\bibitem{Aghanim:2018eyx}
N.~Aghanim et~al.
\newblock {Planck 2018 results. VI. Cosmological parameters}.
\newblock 7 2018.

\bibitem{Riotto:2002yw}
Antonio Riotto.
\newblock {Inflation and the theory of cosmological perturbations}.
\newblock {\em ICTP Lect. Notes Ser.}, 14:317--413, 2003.

\bibitem{Bassett:2005xm}
Bruce~A. Bassett, Shinji Tsujikawa, and David Wands.
\newblock {Inflation dynamics and reheating}.
\newblock {\em Rev. Mod. Phys.}, 78:537--589, 2006.

\bibitem{Sriramkumar:2009kg}
L.~Sriramkumar.
\newblock {An introduction to inflation and cosmological perturbation theory}.
\newblock 4 2009.

\bibitem{Baumann:2009ds}
Daniel Baumann.
\newblock {Inflation}.
\newblock In {\em {Theoretical Advanced Study Institute in Elementary Particle
  Physics}: {Physics of the Large and the Small}}, pages 523--686, 2011.

\bibitem{Akrami:2018odb}
Y.~Akrami et~al.
\newblock {Planck 2018 results. X. Constraints on inflation}.
\newblock 2018.

\bibitem{Martin:2013nzq}
J\'er\^ome Martin, Christophe Ringeval, Roberto Trotta, and Vincent Vennin.
\newblock {The Best Inflationary Models After Planck}.
\newblock {\em JCAP}, 03:039, 2014.

\bibitem{Steinhardetal}
A.~Ijjas, P.~J. Steinhardt, and A.~Loeb.
\newblock Cosmic inflation theory faces challenges.
\newblock {\em Scientific American}, 316(2), 2017.

\bibitem{ResponsetoSteinhardetal}
A.~Guth et~al.
\newblock A cosmic controversy.
\newblock {\em Scientific American}, 316(5), 2017.

\bibitem{Akrami:2019izv}
Y.~Akrami et~al.
\newblock {Planck 2018 results. IX. Constraints on primordial non-Gaussianity}.
\newblock {\em Astron. Astrophys.}, 641:A9, 2020.

\bibitem{Starobinsky:1980te}
Alexei~A. Starobinsky.
\newblock {A New Type of Isotropic Cosmological Models Without Singularity}.
\newblock {\em Phys. Lett. B}, 91:99--102, 1980.

\bibitem{Ashtekar:2006rx}
Abhay Ashtekar, Tomasz Pawlowski, and Parampreet Singh.
\newblock {Quantum nature of the big bang}.
\newblock {\em Phys. Rev. Lett.}, 96:141301, 2006.

\bibitem{Ashtekar:2006wn}
Abhay Ashtekar, Tomasz Pawlowski, and Parampreet Singh.
\newblock {Quantum Nature of the Big Bang: Improved dynamics}.
\newblock {\em Phys. Rev.}, D74:084003, 2006.

\bibitem{Ashtekar:2011ni}
Abhay Ashtekar and Parampreet Singh.
\newblock {Loop Quantum Cosmology: A Status Report}.
\newblock {\em Class. Quant. Grav.}, 28:213001, 2011.

\bibitem{Agullo:2016tjh}
Ivan Agullo and Parampreet Singh.
\newblock {Loop Quantum Cosmology}.
\newblock In Abhay Ashtekar and Jorge Pullin, editors, {\em Loop Quantum
  Gravity: The First 30 Years}, pages 183--240. WSP, 2017.

\bibitem{Agullo:2013dla}
Ivan Agullo and Alejandro Corichi.
\newblock {Loop Quantum Cosmology}.
\newblock In Abhay Ashtekar and Vesselin Petkov, editors, {\em Springer
  Handbook of Spacetime}, pages 809--839. 2014.

\bibitem{Bojowald:2001xe}
Martin Bojowald.
\newblock {Absence of singularity in loop quantum cosmology}.
\newblock {\em Phys. Rev. Lett.}, 86:5227--5230, 2001.

\bibitem{Ashtekar:2003hd}
Abhay Ashtekar, Martin Bojowald, and Jerzy Lewandowski.
\newblock {Mathematical structure of loop quantum cosmology}.
\newblock {\em Adv. Theor. Math. Phys.}, 7(2):233--268, 2003.

\bibitem{MenaMarugan:2011va}
G.~A. Mena~Marugan.
\newblock {Loop Quantum Cosmology: A cosmological theory with a view}.
\newblock {\em J. Phys. Conf. Ser.}, 314:012012, 2011.

\bibitem{Banerjee:2011qu}
Kinjal Banerjee, Gianluca Calcagni, and Mercedes Martin-Benito.
\newblock {Introduction to loop quantum cosmology}.
\newblock {\em SIGMA}, 8:016, 2012.

\bibitem{Bojowald:2008jv}
Martin Bojowald, Golam~Mortuza Hossain, Mikhail Kagan, and
  S.~Shankaranarayanan.
\newblock {Gauge invariant cosmological perturbation equations with corrections
  from loop quantum gravity}.
\newblock {\em Phys. Rev. D}, 79:043505, 2009.
\newblock [Erratum: Phys.Rev.D 82, 109903 (2010)].

\bibitem{Bojowald:2010me}
Martin Bojowald and Gianluca Calcagni.
\newblock {Inflationary observables in loop quantum cosmology}.
\newblock {\em JCAP}, 03:032, 2011.

\bibitem{Agullo:2012sh}
Ivan Agullo, Abhay Ashtekar, and William Nelson.
\newblock {A Quantum Gravity Extension of the Inflationary Scenario}.
\newblock {\em Phys. Rev. Lett.}, 109:251301, 2012.

\bibitem{Agullo:2012fc}
Ivan Agullo, Abhay Ashtekar, and William Nelson.
\newblock {Extension of the quantum theory of cosmological perturbations to the
  Planck era}.
\newblock {\em Phys. Rev. D}, 87(4):043507, 2013.

\bibitem{Agullo:2013ai}
Ivan Agullo, Abhay Ashtekar, and William Nelson.
\newblock {The pre-inflationary dynamics of loop quantum cosmology: Confronting
  quantum gravity with observations}.
\newblock {\em Class. Quant. Grav.}, 30:085014, 2013.

\bibitem{Fernandez-Mendez:2013jqa}
Mikel Fern\'andez-M\'endez, Guillermo~A. Mena~Marug\'an, and Javier Olmedo.
\newblock {Hybrid quantization of an inflationary model: The flat case}.
\newblock {\em Phys. Rev. D}, 88(4):044013, 2013.

\bibitem{Fernandez-Mendez:2014raa}
Mikel Fern\'andez-M\'endez, Guillermo~A. Mena~Marug\'an, and Javier Olmedo.
\newblock {Effective dynamics of scalar perturbations in a flat
  Friedmann-Robertson-Walker spacetime in Loop Quantum Cosmology}.
\newblock {\em Phys. Rev. D}, 89(4):044041, 2014.

\bibitem{Barrau:2014maa}
Aurelien Barrau, Martin Bojowald, Gianluca Calcagni, Julien Grain, and Mikhail
  Kagan.
\newblock {Anomaly-free cosmological perturbations in effective canonical
  quantum gravity}.
\newblock {\em JCAP}, 05:051, 2015.

\bibitem{deBlas:2016puz}
Daniel~Mart{\'\i}n de~Blas and Javier Olmedo.
\newblock {Primordial power spectra for scalar perturbations in loop quantum
  cosmology}.
\newblock {\em JCAP}, 1606(06):029, 2016.

\bibitem{Agullo:2015tca}
Ivan Agullo and Noah~A. Morris.
\newblock {Detailed analysis of the predictions of loop quantum cosmology for
  the primordial power spectra}.
\newblock {\em Phys. Rev.}, D92(12):124040, 2015.

\bibitem{Agullo:2016hap}
Ivan Agullo, Abhay Ashtekar, and Brajesh Gupt.
\newblock {Phenomenology with fluctuating quantum geometries in loop quantum
  cosmology}.
\newblock {\em Class. Quant. Grav.}, 34(7):074003, 2017.

\bibitem{Ashtekar:2016wpi}
Abhay Ashtekar and Brajesh Gupt.
\newblock {Quantum Gravity in the Sky: Interplay between fundamental theory and
  observations}.
\newblock {\em Class. Quant. Grav.}, 34(1):014002, 2017.

\bibitem{Martinez:2016hmn}
Florencia~Ben\'\i{}tez Mart\'\i{}nez and Javier Olmedo.
\newblock {Primordial tensor modes of the early Universe}.
\newblock {\em Phys. Rev. D}, 93(12):124008, 2016.

\bibitem{Gomar:2017yww}
Laura Castell\'o~Gomar, Guillermo~A. Mena~Marug\'an, Daniel Mart\'in De~Blas,
  and Javier Olmedo.
\newblock {Hybrid loop quantum cosmology and predictions for the cosmic
  microwave background}.
\newblock {\em Phys. Rev. D}, 96(10):103528, 2017.

\bibitem{Zhu:2017jew}
Tao Zhu, Anzhong Wang, Gerald Cleaver, Klaus Kirsten, and Qin Sheng.
\newblock {Pre-inflationary universe in loop quantum cosmology}.
\newblock {\em Phys. Rev.}, D96(8):083520, 2017.

\bibitem{Agullo:2018wbf}
Ivan Agullo.
\newblock {Primordial power spectrum from the Dapor-Liegener model of loop
  quantum cosmology}.
\newblock {\em Gen. Rel. Grav.}, 50(7):91, 2018.

\bibitem{Li:2019qzr}
Bao-Fei Li, Parampreet Singh, and Anzhong Wang.
\newblock {Primordial power spectrum from the dressed metric approach in loop
  cosmologies}.
\newblock {\em Phys. Rev. D}, 101(8):086004, 2020.

\bibitem{ElizagaNavascues:2020fai}
Beatriz Elizaga~Navascu\'es, Guillermo A.~Mena Marug\'an, and Santiago Prado.
\newblock {Non-oscillating power spectra in Loop Quantum Cosmology}.
\newblock {\em Class. Quant. Grav.}, 38(3):035001, 2020.

\bibitem{Li:2020mfi}
Bao-Fei Li, Javier Olmedo, Parampreet Singh, and Anzhong Wang.
\newblock {Primordial scalar power spectrum from the hybrid approach in loop
  cosmologies}.
\newblock {\em Phys. Rev. D}, 102:126025, 2020.

\bibitem{Agullo:2020wur}
Ivan Agullo, Javier Olmedo, and V.~Sreenath.
\newblock {Predictions for the Cosmic Microwave Background from an Anisotropic
  Quantum Bounce}.
\newblock {\em Phys. Rev. Lett.}, 124(25):251301, 2020.

\bibitem{Agullo:2020iqv}
Ivan Agullo, Javier Olmedo, and V.~Sreenath.
\newblock {Observational consequences of Bianchi I spacetimes in loop quantum
  cosmology}.
\newblock {\em Phys. Rev. D}, 102(4):043523, 2020.

\bibitem{Ashtekar:2020gec}
Abhay Ashtekar, Brajesh Gupt, Donghui Jeong, and V.~Sreenath.
\newblock {Alleviating the Tension in the Cosmic Microwave Background using
  Planck-Scale Physics}.
\newblock {\em Phys. Rev. Lett.}, 125(5):051302, 2020.

\bibitem{ElizagaNavascues:2020uyf}
Beatriz Elizaga~Navascu\'es and Guillermo A.~Mena Marug\'an.
\newblock {Hybrid Loop Quantum Cosmology: An Overview}.
\newblock 11 2020.

\bibitem{Martin-Benito:2021szh}
Mercedes Mart\'\i{}n-Benito, Rita~B. Neves, and Javier Olmedo.
\newblock {States of Low Energy in bouncing inflationary scenarios in Loop
  Quantum Cosmology}.
\newblock 4 2021.

\bibitem{Li:2021mop}
Bao-Fei Li, Parampreet Singh, and Anzhong Wang.
\newblock {Phenomenological implications of modified loop cosmologies: an
  overview}.
\newblock {\em Front. Astron. Space Sci.}, 8:701417, 2021.

\bibitem{Ashtekar:2021izi}
Abhay Ashtekar, Brajesh Gupt, and V.~Sreenath.
\newblock {Cosmic Tango Between the Very Small and the Very Large: Addressing
  CMB Anomalies Through Loop Quantum Cosmology}.
\newblock {\em Front. Astron. Space Sci.}, 8:76, 2021.

\bibitem{Agullo:2021oqk}
Ivan Agullo, Dimitrios Kranas, and V.~Sreenath.
\newblock {Anomalies in the Cosmic Microwave Background and Their Non-Gaussian
  Origin in Loop Quantum Cosmology}.
\newblock {\em Front. Astron. Space Sci.}, 8:703845, 2021.

\bibitem{Agullo:2017eyh}
Ivan Agullo, Boris Bolliet, and V.~Sreenath.
\newblock {Non-Gaussianity in Loop Quantum Cosmology}.
\newblock {\em Phys. Rev.}, D97(6):066021, 2018.

\bibitem{Agullo:2015aba}
Ivan Agullo.
\newblock {Loop quantum cosmology, non-Gaussianity, and CMB power asymmetry}.
\newblock {\em Phys. Rev.}, D92:064038, 2015.

\bibitem{Sreenath:2019uuo}
Vijayakumar Sreenath, Ivan Agullo, and Boris Bolliet.
\newblock {Computation of non-Gaussianity in loop quantum cosmology}.
\newblock In {\em {15th Marcel Grossmann Meeting on Recent Developments in
  Theoretical and Experimental General Relativity, Astrophysics, and
  Relativistic Field Theories}}, 4 2019.

\bibitem{Zhu:2017onp}
Tao Zhu, Anzhong Wang, Klaus Kirsten, Gerald Cleaver, and Qin Sheng.
\newblock {Primordial non-Gaussianity and power asymmetry with quantum
  gravitational effects in loop quantum cosmology}.
\newblock {\em Phys. Rev. D}, 97(4):043501, 2018.

\bibitem{Aghanim:2019ame}
N.~Aghanim et~al.
\newblock {Planck 2018 results. V. CMB power spectra and likelihoods}.
\newblock 7 2019.

\bibitem{Diener:2013uka}
Peter Diener, Brajesh Gupt, and Parampreet Singh.
\newblock {Chimera: A hybrid approach to numerical loop quantum cosmology}.
\newblock {\em Class. Quant. Grav.}, 31:025013, 2014.

\bibitem{Diener:2014mia}
Peter Diener, Brajesh Gupt, and Parampreet Singh.
\newblock {Numerical simulations of a loop quantum cosmos: robustness of the
  quantum bounce and the validity of effective dynamics}.
\newblock {\em Class. Quant. Grav.}, 31:105015, 2014.

\bibitem{Meissner:2004ju}
Krzysztof~A. Meissner.
\newblock {Black hole entropy in loop quantum gravity}.
\newblock {\em Class. Quant. Grav.}, 21:5245--5252, 2004.

\bibitem{Ashtekar:2009mm}
Abhay Ashtekar and David Sloan.
\newblock {Loop quantum cosmology and slow roll inflation}.
\newblock {\em Phys. Lett.}, B694:108--112, 2011.

\bibitem{Ashtekar:2011rm}
Abhay Ashtekar and David Sloan.
\newblock {Probability of Inflation in Loop Quantum Cosmology}.
\newblock {\em Gen. Rel. Grav.}, 43:3619--3655, 2011.

\bibitem{Bolliet:2017czc}
Boris Bolliet, Aurélien Barrau, Killian Martineau, and Flora Moulin.
\newblock {Some Clarifications on the Duration of Inflation in Loop Quantum
  Cosmology}.
\newblock {\em Class. Quant. Grav.}, 34(14):145003, 2017.

\bibitem{Bonga:2015kaa}
B{\'e}atrice Bonga and Brajesh Gupt.
\newblock {Inflation with the Starobinsky potential in Loop Quantum Cosmology}.
\newblock {\em Gen. Rel. Grav.}, 48(6):71, 2016.

\bibitem{Bonga:2015xna}
B{\'e}atrice Bonga and Brajesh Gupt.
\newblock {Phenomenological investigation of a quantum gravity extension of
  inflation with the Starobinsky potential}.
\newblock {\em Phys. Rev.}, D93(6):063513, 2016.

\bibitem{Zhu:2016dkn}
Tao Zhu, Anzhong Wang, Klaus Kirsten, Gerald Cleaver, and Qin Sheng.
\newblock {Universal features of quantum bounce in loop quantum cosmology}.
\newblock {\em Phys. Lett. B}, 773:196--202, 2017.

\bibitem{Maldacena:2002vr}
Juan~Martin Maldacena.
\newblock {Non-Gaussian features of primordial fluctuations in single field
  inflationary models}.
\newblock {\em JHEP}, 05:013, 2003.

\bibitem{Agullo:2020fbw}
Ivan Agullo, Dimitrios Kranas, and V.~Sreenath.
\newblock {Anomalies in the CMB from a cosmic bounce}.
\newblock {\em Gen. Rel. Grav.}, 53(2):17, 2021.

\bibitem{Agullo:2020cvg}
Ivan Agullo, Dimitrios Kranas, and V.~Sreenath.
\newblock {Large scale anomalies in the CMB and non-Gaussianity in bouncing
  cosmologies}.
\newblock {\em Class. Quant. Grav.}, 38(6):065010, 2021.

\bibitem{Komatsu:2001rj}
Eiichiro Komatsu and David~N. Spergel.
\newblock {Acoustic signatures in the primary microwave background bispectrum}.
\newblock {\em Phys. Rev. D}, 63:063002, 2001.

\bibitem{Bolliet:2015bka}
Boris Bolliet, Julien Grain, Clement Stahl, Linda Linsefors, and Aurelien
  Barrau.
\newblock {Comparison of primordial tensor power spectra from the deformed
  algebra and dressed metric approaches in loop quantum cosmology}.
\newblock {\em Phys. Rev.}, D91(8):084035, 2015.

\bibitem{durrer_2020}
Ruth Durrer.
\newblock {\em The Cosmic Microwave Background}.
\newblock Cambridge University Press, 2 edition, 2020.

\bibitem{Dodelson:2003ft}
Scott Dodelson.
\newblock {\em {Modern Cosmology}}.
\newblock Academic Press, Amsterdam, 2003.

\bibitem{Weinberg_2008}
Steven Weinberg.
\newblock {\em Cosmology}.
\newblock Oxford University Press, 2008.

\bibitem{Fergusson:2006pr}
J.~R. Fergusson and Edward P.~S. Shellard.
\newblock {Primordial non-Gaussianity and the CMB bispectrum}.
\newblock {\em Phys. Rev. D}, 76:083523, 2007.

\bibitem{2010AdAst2010E..73L}
Michele {Liguori}, Emiliano {Sefusatti}, James~R. {Fergusson}, and E.~P.~S.
  {Shellard}.
\newblock {Primordial Non-Gaussianity and Bispectrum Measurements in the Cosmic
  Microwave Background and Large-Scale Structure}.
\newblock {\em Advances in Astronomy}, 2010:980523, January 2010.

\bibitem{2012JCAP...12..032F}
J.~R. {Fergusson}, M.~{Liguori}, and E.~P.~S. {Shellard}.
\newblock {The CMB bispectrum}.
\newblock {\em JCAP}, 2012(12):032, December 2012.

\bibitem{2011JCAP...07..034B}
D.~{Blas}, J.~{Lesgourgues}, and T.~{Tram}.
\newblock {The Cosmic Linear Anisotropy Solving System (CLASS). Part II:
  Approximation schemes}.
\newblock {\em JCAP}, 7:034, July 2011.

\bibitem{Delgado:2021mxu}
Paola C.~M. Delgado, Ruth Durrer, and Nelson Pinto-Neto.
\newblock {The CMB bispectrum from bouncing cosmologies}.
\newblock {\em JCAP}, 11:024, 2021.

\bibitem{vanTent:2022vgy}
Bartjan van Tent, Paola C.~M. Delgado, and Ruth Durrer.
\newblock {Constraining the bispectrum from bouncing cosmologies with Planck}.
\newblock 12 2022.

\bibitem{Chen:2006xjb}
Xingang Chen, Richard Easther, and Eugene~A. Lim.
\newblock {Large Non-Gaussianities in Single Field Inflation}.
\newblock {\em JCAP}, 06:023, 2007.

\bibitem{Munchmeyer:2014nqa}
Moritz M\"unchmeyer, Fran\c{c}ois Bouchet, Mark~G. Jackson, and Benjamin
  Wandelt.
\newblock {The Komatsu Spergel Wandelt estimator for oscillations in the cosmic
  microwave background bispectrum}.
\newblock {\em Astron. Astrophys.}, 570:A94, 2014.

\end{thebibliography}

\end{document}